\documentclass{aa}
\usepackage{graphicx,times}

\def\la{\lower.5ex\hbox{$\; \buildrel < \over \sim \;$}}
\def\ga{\lower.5ex\hbox{$\; \buildrel > \over \sim \;$}}

\begin{document}

   \title{The quenching of star formation in accretion-driven clumpy turbulent tori of active galactic nuclei}

   %\authorrunning{Vollmer \& Davies}
   \titlerunning{The quenching of star formation in accretion-driven turbulent tori}

   \author{B.~Vollmer\inst{1} \& R.I.~Davies\inst{2}}

   \offprints{B.~Vollmer, e-mail: Bernd.Vollmer@astro.unistra.fr}

   \institute{CDS, Observatoire astronomique de Strasbourg, UMR 7550, 
              11, rue de l'universit\'e, 67000 Strasbourg, France \and
              Max Planck Insitut f\"ur extraterrestrische Physik, Postfach 1312,
              85741, Garching, Germany}

   \date{Received / Accepted}

\abstract{
Galactic gas-gas collisions involving  a turbulent multiphase ISM share common ISM properties: 
dense extraplanar gas visible in CO, large linewidths ($\ga 50$~km\,s$^{-1}$),
strong mid-infrared H$_{2}$ line emission, low star formation activity, and strong radio continuum emission. 
Gas-gas collisions can occur in the form of ram pressure stripping caused by the rapid motion of a spiral galaxy within the
intracluster medium, galaxy head-on collisions, compression of the intragroup gas and/or galaxy ISM by an intruder galaxy which flies through
the galaxy group at a high velocity, or external gas accretion on an existing gas torus in a galactic center. 
We suggest that the common theme of all these gas-gas interactions is adiabatic compression
of the ISM leading to an increase of the turbulent velocity dispersion of the gas. The turbulent gas clouds are then overpressured and
star formation is quenched. Within this scenario we developed a  model for turbulent clumpy gas disks where the energy to drive turbulence is supplied
by external infall or the gain of potential energy by radial gas accretion within the disk. The cloud size is determined by the size
of a continuous (C-type) shock propagating in dense molecular clouds with a low ionization fraction at a given velocity dispersion. 
We give expressions for the expected volume and
area filling factors, mass, density, column density, and velocity dispersion of the clouds. The latter is based on 
scaling relations of intermittent turbulence whose open parameters are estimated for the Circumnuclear Disk
in the Galactic Center. The properties of the model gas clouds ($\sim 0.1$~pc, $\sim 100$~M$_{\odot}$, $\Delta v  \ga 6$~km\,s$^{-1}$) and 
the external mass accretion rate necessary for the quenching
of the star formation rate due to adiabatic compression ($\dot{M} \sim 1$-$10$~M$_{\odot}$yr$^{-1}$) are consistent with those derived 
from high-resolution H$_{2}$ $2.12$~$\mu$m line observations. Based on these findings, a scenario for the evolution of gas tori in galactic centers
is proposed and the implications for star formation in the Galactic Center are discussed.
 }

\maketitle

\section{Introduction \label{sec:intro}}

What do the Circumnuclear Disk (CND) in the Galactic Center, a thick obscuring AGN torus, 
the ram-pressure stripped and tidally distorted
Virgo spiral galaxy NGC~4438, the colliding Taffy galaxies, and Stephan's Quintet, all have in common? At first glance all systems
are very different. First of all, the spatial scales and timescales differ enormously. The CND and AGN tori have spatial extents
of about $10$~pc and rotation timescales of $10^{4}$~yr, whereas the relevant scales and timescales in NGC~4438, the Taffy galaxies, and Stephan's Quintet 
are of the order of tens of kpc and $100$~Myr. The common property of all systems is that they are undergoing gas-gas collisions with
high energy injection rates. In these collisions, one gaseous body is the turbulent clumpy multi-phase ISM, 
while the other can be of different mean density and temperature (e.g. ISM, intragroup or intracluster gas):
NGC~4438 is affected by ongoing ram pressure caused by its rapid motion through the Virgo intracluster medium 
(Vollmer et al. 2005, 2009), the Taffy galaxies recently experienced a head-on ISM-ISM collision (Condon et al. 1993, Vollmer et al. 2012), 
and the intragroup gas of the Stephan's Quintet is compressed by a high-velocity intruder galaxy.

NGC~4438 had a tidal interaction $\sim 100$~Myr ago (Combes et al. 1988) probably associated with an ISM-ISM collision (Kenney et al. 1995).
In addition, the galaxy undergoes strong on-going ram pressure stripping leading to extraplanar CO emission and
a characteristic double-line profile in the extraplanar region (Vollmer et al. 2005).
Strong radio continuum emission is associated with the extraplanar gas (Vollmer et al. 2009).
The Taffy system attracted attention through its strong radio synchrotron bridge, a very 
unusual feature.  The bridge is H{\sc i}-rich (Condon et al. 1993) and was subsequently found to be rich in molecular 
gas as well through CO observations (Gao et al. 2003, Braine et al. 2003). The CO lines have particularly large linewidths.
Strong H$_{2}$ emission is associated with the bridge gas (Peterson et al. 2012).
Whereas the ISM-ISM collision of the Taffy galaxies occurred  $\sim 20$~Myr ago, the gas-gas
collision in Stephan's Quintet is on-going and involves an intruder galaxy hitting the intra-group medium of this
compact group with a velocity of $\sim 1000$~km\,s$^{-1}$. The resulting large-scale shock ($\sim 40$~kpc) 
emits in X-ray (Trinchieri et al. 2003) and radio continuum emission (van der Hulst 1981).
Powerful high-velocity dispersion molecular hydrogen is associated with the intergalactic shock wave
(Appleton et al. 2006, Guillard et al. 2009). About $5 \times 10^{8}$~M$_{\odot}$ of warm 
H$_{2}$ spread over $\sim 480$~kpc$^{2}$ were found in the main shock region (Cluver et al. 2010).
In addition, CO(1--0), (2--1) and (3--2) line emission have been detected in this region with complex profiles, 
spanning a velocity range $\sim 1000$~km\,s$^{-1}$ (Guillard et al. 2012).
The intra-group material involved has most probably been tidally stripped by a past galaxy--galaxy interaction
(Renaud et al. 2010, Hwang et al. 2012).

All these regions outside the galactic disks formed by galactic gas-gas collisions involving high energy injection rates, have high gas densities (CO emission), 
involve huge molecular gas masses, show large linewidths, and have strong associated H$_{2}$ and radio continuum emission.
In addition, star formation is significantly reduced in all these regions (NGC~4438: Vollmer et al. 2009; Taffy galaxies: Vollmer et al. 2012;
Stephan's Quintet: Guillard et al. 2012).
Gas tori in active galactic centers have many of these properties in common with these regions: they involve huge gas masses
(Hicks et al. 2009), display high gas densities and large linewidths (Davies et al. 2007, Sani et al. 2012), and have
a significantly reduced star formation rate, i.e. they are classified as post-starburst regions (Davies et al. 2007).
Based on these similarities, we suggest that the thick, obscuring AGN gas tori undergo a violent ISM-ISM collision
caused by infalling material which hits the torus. The mechanical energy input drives the gas turbulence and quenches star formation
as in galaxy-scale ISM-ISM collisions.

The ISM in galaxies and galactic center gas tori is clumpy. Thermal instabilities triggered by small-scale gas compression
lead to the condensation of dense cold clouds (Parravano 1987). If the turbulent crossing time is smaller than the free-fall
time, the clouds become selfgravitating and might collapse.
The CND in the Galactic Center is made of gas clouds with masses
of a few $10$~M$_{\odot}$, an area filling factor of $\Phi_{\rm A} \sim 0.1$ and a volume filling factor of $\Phi_{\rm V} \sim 0.01$
(Jackson et al. 1993; however, Requena-Torres et al. 2012 suggest a much higher volume filling factor of $\Phi_{\rm V} \sim 0.2$). 
Since the densities derived from the HCN lines (Jackson et al. 1993) are close to those of selfgravitating clouds with a thermal sound 
speed of $\sim 1$~km\,s$^{-1}$, Vollmer \& Duschl (2001a) and Vollmer et al. (2004) proposed a model of a collisional disk made of stable
gas clouds. This model was extended to AGN gas tori in Vollmer et al. (2008). These authors suggested the following scenario for
the evolution of an AGN gas torus: during a rapid massive infall\footnote{The infall has to be rapid to avoid significant star formation on 
the way to the central $10$~pc and massive to insure a torus gas mass of $\sim 10^{6}$~M$_{\odot}$.}, 
gas accumulates at a distance of $\sim 10$~pc around the central black hole (Davies et al. 2007, Lira et al. 2013).
At this stage the torus forms stars giving rise to a nuclear starburst. After $\sim 10$~Myr the SN explosions remove the torus intercloud
gas leaving a torus made of dense clumps. Due to a high mass accretion rate, the torus has a high velocity dispersion, i.e. it is thick.
Its radial mass accretion is high enough to power the central AGN.
This collisional torus has a rather low star formation rate. Once the external mass accretion rate decreases the collisional torus
eventually becomes thin.

In the present article, we modify this scenario by replacing the collisional torus (Sect.~\ref{sec:theory}) by a turbulent clumpy torus
(Sect.~\ref{sec:turbulent}). We show that both types of models lead to gas clouds of comparable masses and sizes (Sect.~\ref{sec:application}).
However, the turbulent torus clouds have a much higher velocity dispersion ($\geq 6$~km\,s$^{-1}$) than the collisional CND clouds
($\geq 1$~km\,s$^{-1}$). Since the collisional clouds are by definition selfgravitating, the torus clouds are overpressured
with respect to their turbulent pressure.
This prevents gravitational collapse and quenches star formation (Sect.~\ref{sec:quench}). The model cloud properties are compared
to those derived from subarcsecond molecular gas emission line observations of two Seyfert galaxies in Sect.~\ref{sec:seyfert}.
The implications for the recent star formation history in the Galactic Center are discussed in Sect.~\ref{sec:sfrgc} and
we give our conclusions in Sect.~\ref{sec:conclusions}.

\section{The theory of clumpy gas disks \label{sec:theory}}

Galactic accretion disks generally are clumpy: gas disks of spiral galaxies contain gas clumps of cold neutral
hydrogen (e.g., Strasser et al. 2007) with embedded giant molecular clouds (e.g., Solomon et al. 1987);
gas tori around central galactic black holes also contain clumps of high volume densities (e.g., Krolik \& Begelman 1988, 
G\"{u}sten et al. 1987). The formation of regions of overdense gas is caused by thermal instabilities and selfgravity.
In turbulent galactic disks, gas clumps are of transient nature with lifetimes of about a crossing time (e.g., Dobbs \& Pringle 2013).
The governing gas physics of such disks are highly time-dependent and intrinsically stochastic. 
Over a long-enough timescale, turbulent motion of clumps is expected to redistribute angular momentum 
in the gas disk like an effective viscosity would do. This allows accretion of gas towards the center and makes it possible to treat 
the disk as an accretion disk (e.g., Pringle 1981). This gaseous turbulent accretion disk rotates in a given gravitational
potential $\Phi$ with an angular velocity $\Omega=\sqrt{R^{-1}\frac{{\rm d}\Phi}{{\rm d}R}}$, where $R$ is the disk radius. The disk has an
effective turbulent viscosity that is responsible for mass accretion and outward angular momentum transport. 
The turbulent velocity can be driven by SN explosions or external gas infall, which stir the disk and lead to viscous transport of
angular momentum. In addition, star formation might remove gas from the viscous evolution.
Following Lin \& Pringle (1987), the evolution of the gas surface density is given by
\begin{equation}
\frac{\partial \Sigma}{\partial t}=-\frac{1}{R}\frac{\partial}{\partial R}\left(
\frac{(\partial/\partial R)[\nu \Sigma R^3 ({\rm d}\Omega/{\rm }dR)]}{({\rm d}/{\rm d}R)(R^2 \Omega)}\right)
-\dot{\Sigma}_{*}+\dot{\Sigma}_{\rm sph}\ ,
\label{eq:linpringle}
\end{equation}
where $\nu$ is the gas disk viscosity, $\Omega$ the angular velocity. Gas which does not accrete within the disk plane gives rise
to the spherical mass accretion rate $\dot{\Sigma}_{\rm sph}$. 
In contrast to Lin \& Pringle (1987) we assume a continuous and non-zero spherical gas mass 
accretion rate. 
By approximating $\partial/\partial R \sim 1/R$, the global viscous evolution becomes
\begin{equation}
\frac{\partial \Sigma}{\partial t} \sim \frac{\Sigma \nu}{R^2}-\dot{\Sigma}_{*}+\dot{\Sigma}_{\rm sph}\ .
\label{eq:parsigma}
\end{equation}
Radial integration of Eq.~\ref{eq:parsigma} then gives
\begin{equation}
\frac{\partial M}{\partial t} \sim \dot{M}-\dot{M}_{*}+\dot{M}_{\rm sph}\ ,
\end{equation}
where $M$ is the disk gas mass and $\dot{M}$ the radial disk accretion rate. 
The external mass accretion rate is thus $\dot{M}_{\rm ext}=\dot{M}+\dot{M}_{\rm sph}$.
We note that in thick gas disks or tori, the disk occupies a significant solid angle and $\dot{M}_{\rm sph}$ decreases
in favor of $\dot{M}$.

If, in thin starforming galactic disks, the external mass accretion rate keeps the combined Toomre parameter of the gas and stars 
smoothed over a few rotation periods constant, the gas surface density will only vary slowly with changes in the dark halo
mass distribution (via $\Omega$) and the stellar disk structure, the gas loss due to star formation is balanced by spherical accretion
as suggested by Fraternali et al. (2008) and Marinacci et al (2010),
and the gas disk can be regarded as being stationary\footnote{Indeed, local spiral galaxies show a $Q_{\rm tot}$ not too far away 
from unity ($1.3$-$2.5$; Leroy et al. 2008).} ($\partial M / \partial t = 0$).

In the absence of star formation in thick gas tori around central galactic black holes (see Sect.~\ref{sec:quench}), the disk can be regarded as stationary if
(i) the external infall mainly occurs within the disk plane, i.e. $\dot{M}_{\rm sph} \ll \dot{M}$ or $\dot{M}_{\rm ext} \sim \dot{M}$,  and (ii)
over a sufficiently large timescale, the time-averaged external mass infall is constant.
Condition (i) is quite naturally fulfilled for a thick gas disk/torus which is mainly fed at the outer edge. The relevant timescale for
disk evolution is then the viscous timescale. Condition (ii) is a quite strong assumption which might not be
fulfilled in all cases. In a recent numerical work, Gaspari et al. (2013) studied the behavior of cold accretion down to the central part of an elliptical galaxy.
They showed that in a cooling, heated and turbulent atmosphere of
elliptical galaxies, thermal instabilities become quickly nonlinear, leading to the condensation of dense cold clouds in a 
turbulent medium, up to several kpc from the center. 
The clouds then fall toward the very inner region and strongly collide within $10$-$100$~pc. 
The inelastic collisions promote angular momentum cancellation, strongly boosting black hole accretion.
Thus, the collision of gas clumps can dominate the distribution of angular momentum and thus gas accretion.
Within this scenario, the external mass accretion rate is highly time-dependent and intrinsically stochastic (Fig.~3 and 5 of Gaspari et al. 2013).
For our purpose, one has to time-average the external mass accretion rate on scales of $\Delta t \sim 5$~Myr, the typical timescale of the fluctuations
of the mass accretion rate. This timescale is comparable to the viscous timescale of a thick torus ($t_{\rm visc}=R^{2}/\nu$; see Eq.~\ref{eq:tvisc}).

For a stationary gas disk which fulfills conditions (i) and (ii) the local mass and momentum conservations yield:
\begin{equation}
\nu \Sigma=\frac{\dot{M}}{2\pi}\ ,
\label{eq:transport}
\end{equation}
where $\dot{M}$ is the radial mass accretion rate within the disk plane.

In this work we are not interested in the detailed radial distributions of the disk properties, but look at the integrated
values. To measure the gas content of the gas disk, we use the Toomre parameter $Q$ of the gas
\begin{equation}
Q=\frac{v_{\rm turb}}{v_{\rm rot}} \frac{M_{\rm dyn}}{M}\ ,
\end{equation}
where $v_{\rm turb}$ and $v_{\rm rot}$ are the turbulent and rotation velocities, and $M_{\rm dyn}$ is the total enclosed mass.
Hicks et al. (2009) and Sani et al. (2012) found  $Q > 1$ in AGN gas tori where star formation is quenched.

The present rather simple analytical model of steady clumpy accretion disks aims at capturing the basic gas
physics of (i) turbulent and (ii) collisional disks within the framework set by Vollmer \& Beckert (2002, 2003) and Vollmer et al. (2004).
In case (i) the ISM is regarded as
a single entity which changes phase (molecular, atomic, ionized) according 
to internal (gas density, pressure, magnetic field) and external (gravitation,
radiation field, winds) conditions. Energy is injected into a turbulent cascade 
at the driving lengthscale (large scale) and dissipated 
at the dissipation lengthscale (small scale). In Vollmer \& Beckert (2002, 2003) we identified the dissipation 
lengthscale with the characteristic size of selfgravitating clouds in starforming disks. 
These clouds decouple from the the turbulent cascade and constitute the first energy sink.
The source of energy which is injected at the driving scale to maintain turbulence
can be either (i) mass accretion in the gravitational
potential of the galactic center ({\it fully gravitational FG model})
or (ii) supernova explosions ({\it SN model}). AGN feedback, which represents an additional energy source for turbulence is not included in the model.
In the collisional case energy is also supplied in the process of mass accretion in the gravitational
potential of the galactic center and dissipated via partially
inelastic cloud--cloud collisions. The actual dissipation rate in individual collisions is largely unknown.
The disk evolution is mainly driven by the external mass accretion rate.
Since these models are equilibrium models, we assume that the mass accretion rate
is constant throughout the region of interest when averaged for a sufficiently long time 
($\sim \Omega^{-1}$; here $\Omega$ is the angular velocity of circular orbits in the gravitational 
potential of the galactic nucleus) and maintained for at least the turnover timescale $R/v_{\rm turb}$ 
of gas in the disk, where $R$ is the distance from the center of the galaxy and $v_{\rm turb}$ the 
characteristic speed of turbulent eddies.
All models provide the global parameters of the disk 
and the local parameters of the most massive clouds (see Table~\ref{tab:model}).
The free parameters of the models are the Toomre parameter $Q$, the mass accretion rate $\dot{M}$,
the disk transparency $\vartheta$ (Eq.~\ref{eq:tau}) for the collisional model, and the
scaling parameter between the driving lengthscale and the cloud size in the turbulent model. 
Other parameters are fixed using the Galactic values (Vollmer \& Beckert 2002, 2003). Each model has an associated star formation rate.
In the following we describe these models in more detail.

In the present article we extend this theory to non-starforming clumpy accretion disks
where the source of energy to maintain turbulence is external gas infall (mechanical energy) and subsequent mass accretion in the gravitational
potential of the galactic center. We assume that the external mass infall occurs mainly within the disk plane. The disk mass accretion rate
is thus equivalent to the external mass accretion rate. The turbulent energy is mainly dissipated by molecular line radiation 
(CO, H$_{2}$, H$_{2}$O; see, e.g., Neufeld \& Kaufman 1993) from the dense clouds. In contrast to the previous work, dense clouds do not need to be
selfgravitating.

\section{Collisional disks \label{sec:collisional}}

The disk clouds are supposed to be stable (via selfgravitation and magnetic fields) and long-lived (several rotation periods). 
In this case their collisions will give rise to angular 
momentum redistribution which can be described by an effective viscosity. 
An equilibrium disk can be formed if there are fragmenting
collisions or partially inelastic collisions (the clouds are supposed to be magnetized).
It follows that the collisional energy dissipation rate can be written as
\begin{equation}
\frac{\Delta E}{\Delta A\,\Delta t}=f \frac{\Sigma v_{\rm turb}^{2}}{t_{\rm coll}}=
f \frac{\Sigma v_{\rm turb}^{3}}{l_{\rm coll}}=f \frac{\Sigma v_{\rm turb}^{3}}{\vartheta H}\ ,
\label{eq:eatcoll}
\end{equation}
where $\Sigma= 2 \rho H$ is the mean gas surface density, $\rho$ the mean gas density, $H$ the disk height,
$v_{\rm turb}$ the turbulent velocity, $t_{\rm coll}$ the collision timescale, $l_{\rm coll}$ the collision lengthscale,
and $\vartheta$ the disk transparency. The disk mean density is linked to the cloud density via the
volume filling factor $\rho=\rho_{\rm cl} \Phi_{\rm V}$.

The factor $f$ accounts for the mean fraction 
of cloud mass participating in the highly supersonic cloud collisions. For constant density 
clouds Krolik \& Begelman (1988) argue that $f = 0.2$. For more centrally condensed, 
self-gravitating clouds we will use a factor $f=0.1$ as in Vollmer et al. (2008). 
If the collisional timescale $t_{\rm coll}$ is longer or
equal to the dynamical timescale, the resulting viscosity can be written as 
\begin{equation} \label{eq:vis}
\nu = f \vartheta^{-1} v_{\rm turb} H\ ,
\label{eq:tau}
\end{equation}
where the disk transparency $\vartheta = t_{\rm coll}\Omega > 1$ and $H$ is the disk height.

The cloud size $r_{\rm cl}$ and the volume filling factor $\Phi_{\rm V}$ of clouds 
can be derived using their mean free path (see Vollmer et al. 2004) 
\begin{equation}
l_{\rm coll}=\vartheta H = \frac{4r_{\rm cl}}{3 \Phi_{\rm V}}
\end{equation}
and the fact that the clouds are selfgravitating
\begin{equation}
t^{\rm cl}_{\rm ff}=\sqrt{\frac{3 \pi \Phi_{\rm V}}{32 G \rho}}= t_{\rm s} = \frac{r_{\rm cl}}{c_{\rm s}}\ ,
\label{eq:tfftturb}
\end{equation}
where $t^{\rm cl}_{\rm ff}$ is the free fall timescale within clouds, $\rho=\Omega^2/\pi/G/Q$ the disk overall gas density, 
$t_{\rm s}$ the sound crossing timescale, $c_{\rm s}$ the sound speed, and $G$ the gravitation constant.
The area filling factor of the clouds is then
\begin{equation}
\Phi_{\rm A}=\Phi_{\rm V} H/r_{\rm cl}=\frac{4}{3} \vartheta^{-1}\ .
\label{eq:phia}
\end{equation}
Eq.~\ref{eq:tfftturb}, Eq.~\ref{eq:phia}, and $H=v_{\rm turb}/\Omega$ lead to
\begin{equation}
r_{\rm cl}=\frac{3 \pi^2}{32}\frac{Q c_{\rm s}^2 \Phi_{\rm A}}{\Omega v_{\rm turb}}
\end{equation}
and
\begin{equation}
\Phi_{\rm V}=\frac{3 \pi^2}{32} \frac{Q c_{\rm s}^2 \Phi_{\rm A}^{2}}{v_{\rm turb}^2}\ .
\end{equation}
The cloud mass is then
\begin{equation}
M_{\rm cl}=\frac{4 \pi}{3} \Phi_{\rm V}^{-1} \rho r_{\rm cl}^3=\frac{3 \pi^{4}}{256} \frac{Q c_{\rm s}^{4} \Phi_{\rm A}}{\Omega G v_{\rm turb}}\ .
\label{eq:cloudmass}
\end{equation}

Based on the results of Vollmer et al. (2004), Vollmer et al. (2008) assumed that the cloud mass of all disks is close to that of the 
CND (Jackson et al. 1993), $M_{\rm cl}=10$~M$_{\odot}$.
This is equivalent to a common size, density, and column density of all clouds 
\begin{equation}
r_{\rm cl}=\frac{8}{\pi^{2}} \frac{M_{\rm cl} G}{c_{\rm s}^{2}}\ ,
\label{eq:rcoll}
\end{equation}
\begin{equation}
\rho_{\rm cl}=\frac{3 \pi^{5}}{2048}\frac{c_{\rm s}^{6}}{M_{\rm cl}^{2} G^{3}}\ ,\ {\rm and}
\label{eq:rhocoll}
\end{equation}
\begin{equation}
N_{\rm cl}=\frac{3 \pi^{3}}{256}\frac{c_{\rm s}^{4}}{M_{\rm cl} G^{2}}\ .
\label{eq:Ncoll}
\end{equation}
Assuming a sound speed of $c_{\rm s}=1.5$~km\,s$^{-1}$ leads to $r_{\rm cl}=0.02$~pc, $\rho_{\rm cl}=10^{7}$~cm$^{-3}$, and 
$N_{\rm cl} = 5 \times 10^{23}$~cm$^{-2}$.
The value for the sound speed corresponds to a cloud temperature of $\sim$500~K (Krips et al. 2011 found $T \ge 200$~K for NGC~1068,  
Hailey-Dunsheath et al. 2012 derived $T \sim 600$~K for NGC~1068, Davies et al. 2012 favor $T \sim 300$~K for NGC~3227) when only thermal gas 
pressure is considered. The sound speed is a measure of the pressure support against self-gravity and additional contributions 
to the pressure 
gradient inside clouds like magnetic fields may contribute. With $Q=5$ (Vollmer et al. 2008, Hicks et al. 2009) and $\Omega=100$~km\,s$^{-1}/10$~pc, 
we obtain for the volume filling factor
\begin{equation}
\Phi_{\rm V}=\frac{2048}{3 \pi^{6}}\frac{M_{\rm cl}^{2} G^{2}}{Q c_{\rm s}^{6}}\Omega^{2}=0.003\ .
\label{eq:phivcoll}
\end{equation}
The area filling factor is then
\begin{equation}
\Phi_{\rm A}=\Phi_{\rm V} \frac{H}{r_{\rm cl}} = \frac{256}{3 \pi^{4}} \frac{M_{\rm cl} \Omega G v_{\rm turb}}{Q c_{\rm s}^{4}}\ .
\label{eq:phiacoll}
\end{equation}
Assuming a thick disk with a height of $H=5$~pc ($v_{\rm turb}=H \Omega=50$~km\,s$^{-1}$) yields an area filling factor of $\Phi_{\rm A}=0.8$, i.e.
the disk obscures the central engine.

Since the clouds are assumed to be stable and long-lived, they have to resist tidal shear $\rho_{\rm cl} \ge \Omega^{2}/\pi/G$.
This translates into a minimum sound speed at the inner edge of the torus where tidal shear is maximum of
\begin{equation}
c_{\rm s} \ge \frac{2.97}{\pi} \big(\Omega G M_{\rm cl} \big)^{\frac{1}{3}}=0.73~{\rm km\,s}^{-1}\ .
\end{equation}

\begin{table}
      \caption{Model parameters and their meaning}
         \label{tab:model}
      \[
       \begin{array}{|l|l|}
        \hline
	{\rm large\ scale} & {\rm disk} \\
	\hline
	\hline
	R & {\rm galactic\ radius} \\
	v_{\rm rot} & {\rm rotation\ velocity} \\
	\Omega & {\rm angular\ velocity} \\
	v_{\rm A,0} & {\rm Alfven\ velocity} \\
	M_{\rm dyn} & {\rm total\ enclosed\ (dynamical)\ mass} \\
	M_{\rm gas} & {\rm total\ gas\ mass} \\
	v_{\rm turb} & {\rm gas\ turbulent\ velocity\ dispersion} \\
	Q & {\rm Toomre\ parameter} \\
	\rho & {\rm midplane\ density} \\
	\Sigma & {\rm surface\ density} \\
	H & {\rm disk\ height} \\
	\dot{M} & {\rm disk\ mass\ accretion\ rate} \\
	\dot{M}_{\rm ext} & {\rm external\ mass\ accretion\ rate} \\
	\dot{M}_* & {\rm star\ formation\ rate} \\
	\nu & {\rm gas\ viscosity} \\
	l_{\rm driv} & {\rm turbulent\ driving\ lengthscale} \\
	\Phi_{\rm V} & {\rm cloud\ volume\ filling\ factor} \\
	\Phi_{\rm A} & {\rm cloud\ area\ filling\ factor} \\
	\zeta & {\rm viscosity\ scaling\ parameter} \\
	f & {\rm mean\ fraction\ of\ cloud\ mass} \\
	 & {\rm participating\ in\ collisions} \\
	\vartheta & {\rm disk\ transparency} \\
	\alpha & {\rm collision\ coefficient} \\
	x_{i} & {\rm ionization\ fraction} \\
	\gamma & {\rm ion-neutral\ coupling\ coefficient} \\
	\zeta_{\rm CR} & {\rm cosmic\ ray\ ionization\ rate} \\ 
	L/M & {\rm luminosity\ to\ mass\ ratio} \\
	\lambda & {\rm compression\ lengthscale} \\
	v_{\rm comp} & {\rm compression\ velocity} \\
	\hline
	\hline
	{\rm small\ scale} & {\rm clouds}  \\
	\hline
	\hline
	l & {\rm size\ of \ the \ C-shock} \\
	l_{\rm coll} & {\rm cloud\ mean\ free\ path} \\ 
	M_{\rm cl} & {\rm cloud\ mass} \\
	r_{\rm cl} & {\rm cloud\ radius} \\
	l_{\rm cl} & {\rm cloud\ size} \\
	\rho_{\rm cl} & {\rm cloud\ density} \\
	N_{\rm cl} & {\rm cloud\ surface\ density} \\
	c_{\rm s} & {\rm local\ sound\ speed\ within\ the\ clouds}\\
	t_{\rm ff}^{\rm cl} & {\rm cloud\ free\ fall\ time}\\
	T & {\rm temperature} \\
	\hline
        \end{array}
      \]
\end{table}

\section{Turbulent disks \label{sec:turbulent}}

In contrast to the previous models of turbulent clumpy accretion disks (Vollmer \& Beckert 2002, 2003;
Vollmer et al. 2004, Vollmer et al. 2008), we do not assume that the clouds are selfgravitating.
We thus have to estimate the cloud size in a different way. As in the previous models, the disk evolution is 
mainly driven by the external mass accretion rate.
Since these models are equilibrium models, we assume that the mass accretion rate
is constant throughout the region of interest when averaged and maintained for a sufficiently long time 
($t_{\rm visc} =R^{2}/\nu \sim 5$~Myr). External accretion onto an existing gas disk
leads to shocks in the dense magnetized molecular gas. 
In such a dense gas ($n \ga 10^{4}$~cm$^{-3}$) with low ionization fraction ($x_{i} \la 10^{-6}$) the ionic 
Alfv\'en speed is higher than the shock speed so that ions and magnetic fields smoothly 
transit between upstream and downstream conditions without discontinuities. 
As a result of the ion-neutral drag forces, the transition in the neutrals is also modified and, in the presence of significant cooling, 
all physical quantities vary smoothly in the shock region, 
forming a continuous (C-type) shock (Draine 1980). The combination of low ionization fraction,
strong cooling, and significant (but not high) magnetic fields thus results in a shock in which ion-neutral drag provides
the viscosity, cooling keeps the gas supersonic and the field via the ions provides an extended cushioning layer which
inhibits molecular dissociation (Draine \& McKee 1993). For high shock velocities ($> 50$~km\,s$^{-1}$ at $n=10^{5}$~cm$^{-3}$) molecular
hydrogen is dissociated in the shock layer due to the high temperature and the shock becomes J-type (Le Bourlot et al. 2002). 
Strong multifluid (neutral and ionized medium, magnetic field) J-type shocks behave like classical single fluid shocks.
These shocks have magnetic/ionic precursors in which ion-neutral drift preheats the incoming gas before the shock itself.
For the considered gas densities $n \le 10^{5}$ and shock velocities $\la 60$~km\,s$^{-1}$, the shocks are mainly C-type.
Extreme shock velocities might lead to J-type shocks with magnetic precursors. The compression ratio and thickness of these shocks
are assumed to be close to those of C-type shocks.

Within the strong shock approximation the gas compression rate in the shock is given by
\begin{equation}
r_{\rm sh}=\frac{\rho_{\rm cl}}{\rho}=\sqrt{2}\frac{v_{\rm turb}}{v_{\rm A,0}}\ ,
\end{equation}
where the $v_{\rm A,0}=B/\sqrt{4 \pi \rho}$ is the neutral Alfv\'{e}n speed in the diffuse medium with
magnetic field $B$ (see, e.g. Chen \& Ostriker 2012). The neutral Alfv\'{e}n speed of the ISM clouds of different 
densities is almost constant $v_{\rm A,0} \sim 1$~km\,s$^{-1}$ (Crutcher 1999). 
Assuming recombination-ionization equilibrium, the shock thickness is given by Eq. 43 of Chen \& Ostriker (2012):
\begin{equation}
l=2^{\frac{7}{4}}\frac{\sqrt{v_{\rm A,0} v_{\rm turb}}}{\alpha x_{i} (\mu_{i}/\mu_{n}) \rho}\ ,
\label{eq:lll}
\end{equation}
where $\alpha=3.7 \times 10^{13}$~cm$^{3}$s$^{-1}$g$^{-1}$ (Draine et al. 1983) is the collision coefficient,
$x_{i}$ the ionization fraction, and $(\mu_{i}/\mu_{n})=(30/2.3)=13$ the fraction between the mean ion and neutral molecular weights.
If the so-called frozen-in condition ($n_{i}v_{i}=const$, where $n_{i}$ and $v_{i}$
are the ionized gas density and velocity; e.g. Wardle 1990) is used instead of recombination-ionization equilibrium, Eq.~\ref{eq:lll} 
and the following equations change. This case is developed in Appendix A. 

The ionization fraction is given by 
\begin{equation}
x_{i}= \gamma \big(\frac{\zeta_{\rm CR}}{n_{\rm H}}\big)^{\frac{1}{2}}\ ,
\end{equation}
where $\gamma=600$-$2000$~cm$^{-\frac{3}{2}}$s$^{\frac{1}{2}}$ (Williams et al. 1998; McKee et al. 2010)
and $n_{\rm H}=\rho/(2.3 \times m_{\rm p})$.
For the cosmic ray ionization rate we take the value for the Galactic Center region of
$\zeta_{\rm CR}=2 \times 10^{-15}$~s$^{-1}$ (Yusef-Zadeh et al. 2013).
With $\gamma=600$~cm$^{-\frac{3}{2}}$s$^{\frac{1}{2}}$ the size of the C-shock is
\begin{equation}
l=0.1 \sqrt{\frac{{(v_{\rm A,0}/1~{\rm km\,s}^{-1})\,(v_{\rm turb}/30~{\rm km\,s}^{-1})}}{(n_{\rm H}/5 \times 10^{4}~{\rm cm}^{-3})}}~{\rm pc}\ .
\label{eq:clumpsize}
\end{equation}
The ionization-recombination approximation is justified, because the ionization and recombination timescales ($t_{i} \sim x_{i}/\zeta_{\rm CR}$ 
and $t_{\rm rec} \sim \gamma^{2} n_{i}$) are very short (years) compared to shock crossing timescale ($t_{\rm cross} = l_{\rm cl}/v_{\rm turb}$).

Within the shock the magnetic field is compressed in the direction of shock propagation.
The shock might become unstable and converging gas flows will form within the shock plane.
In this way gas clouds are created which do not need to become selfgravitating.
Within a simplified picture a converging flow bounded by C-shocks has a breadth at least
twice the shock size (Chen \& Ostriker 2012) which we identify with the cloud size:
$l_{\rm cl}=2 \times l$. An approximation for the volume filling factor is 
\begin{equation}
\begin{array}{ll}
\Phi_{\rm V}=&\frac{1}{\sqrt{2}}\frac{v_{\rm A,0}}{v_{\rm turb}}= \\
&0.02 (v_{\rm A,0}/1~{\rm km\,s}^{-1})/(v_{\rm turb}/30~{\rm km\,s}^{-1})\ .
\end{array}
\label{eq:phivturb}
\end{equation}
With the mean density of $\rho=\Omega^2/(\pi G Q)$ the cloud size becomes
\begin{equation}
\begin{array}{ll}
l= & 0.1 (\Omega/1 \times 10^{-5}~{\rm yr}^{-1})^{-1} \times \\
 & \sqrt{(Q/5)\,(v_{\rm A,0}/1~{\rm km\,s}^{-1})\,(v_{\rm turb}/30~{\rm km\,s}^{-1})}~{\rm pc}\\
\end{array}
\label{eq:lturb}
\end{equation}
and the area filling factor is
\begin{equation}
\Phi_{\rm A}=  11.6 \sqrt{\frac{v_{\rm A,0}}{Q v_{\rm turb}}}=0.94 \ ,
\label{eq:phiaturb}
\end{equation}
the mass of the clouds 
\begin{equation}
\begin{array}{l}
M_{\rm cl}=\frac{4}{3}\pi \Phi_{\rm V}^{-1} \rho \big(\frac{l}{2}\big)^{3}=23\,(\Omega/1.5 \times 10^{-5}~{\rm yr}^{-1})^{-1} \times \\
\big((Q/5)(v_{\rm A,0}/1~{\rm km\,s}^{-1})\big)^{\frac{1}{2}} \big(v_{\rm turb}/30~{\rm km\,s}^{-1}\big)^{\frac{5}{2}}~{\rm M}_{\odot}\ ,
\end{array}
\label{eq:mcl}
\end{equation}
and the cloud surface density 
\begin{equation}
\begin{array}{l}
N_{\rm cl}=7 \times 10^{23}\,(\Omega/1.5 \times 10^{-5}~{\rm yr}^{-1}) \times \\
\big((Q/5)(v_{\rm A,0}/1~{\rm km\,s}^{-1})\big)^{-\frac{1}{2}} \big(v_{\rm turb}/30~{\rm km\,s}^{-1}\big)^{\frac{3}{2}}~{\rm cm}^{-2}\ .
\end{array}
\label{eq:Nturb}
\end{equation}

The gas within the clouds is heated via dissipation and the heating is $\Gamma=\rho_{\rm cl} v_{\rm turb,cl}^{3}/l$ with $l=2 r_{\rm cl}$.
The gas is cooled via molecular line radiation and the cooling is $\Lambda=\mathcal{L} n_{\rm cl} n(M)$, where $n(M)$ is the particle
density of the coolant (CO, H$_{2}$, or H$_{2}$O) and the cooling
rate coefficient $\mathcal{L}$ has the units of erg\,cm$^{3}$\,s$^{-1}$. The equilibrium between heating and cooling
sets the gas temperature in the cloud. 

One of the most striking (and least understood) features of turbulence is its intermittent spatial and temporal behavior
(Balkovsky \& Falkovich 1998). 
The structures that arise in a turbulent flow manifest themselves as high peaks at random places and at random times. 
This is reflected in the probability distribution functions of dynamical variables or passively advected scalars. 
They are sensitive measures of deviations from Gaussian statistics. 
Rare strong fluctuations are responsible for extended tails, whereas the much larger regions of low intensity contribute to the peak of the PDF near zero
(Balkovsky \& Falkovich 1998).
In the case of incompressible turbulence the rate of transfer of energy per unit mass in turbulent eddies of size $l$ is constant
$\epsilon = v_{{\rm turb,} l}^{3}/l = const$ (Kolmogorov 1941). Within the framework of the $\beta$-model for intermittent turbulence,
where the turbulent eddies become less and less space filling, the conserved quantity is 
$\bar{\epsilon} = (l_{\rm driv}/l)^{D-3} v_{{\rm turb,} l}^{3}/l = const$ (Frisch et al. 1978), where $D$ is the fractal dimension. 
For $D=3$ the Kolmogorov turbulence is recovered.
 
In the case of compressible turbulence the mean volume rate of energy transfer becomes $\epsilon_{\rm V}= \rho_{l} \epsilon = \rho_{l} v_{{\rm turb,} l}^{3}/l$
(e.g., Fleck 1996). If the density scales with $\rho_{l} \propto l^{-3 \alpha}$, the turbulent velocity dispersion obeys the relation
$v_{l} \propto l^{1/3+\alpha}$ (Fleck 1996, Kritsuk et al. 2007). Extending the $\beta$-model to compressible turbulence leads to 
\begin{equation}
\bar{\epsilon}_{\rm V} = \rho_{l} (l_{\rm driv}/l)^{D-3} v_{{\rm turb,} l}^{3}/l = const\ .
\label{eq:scaling}
\end{equation}
With $\rho_{l} \propto l^{-3 \alpha}$, the turbulent velocity dispersion obeys the relation $v_{l} \propto l^{\frac{1}{3}(3\alpha+D-2)}$.
For $D=2$ (see Sect.~\ref{sec:gc}) one obtains $v_{l} \propto l^{\alpha}$ and the energy flux 
$\Delta E/(\Delta A \Delta t) = \rho_{l} v_{{\rm turb,} l}^{3}$ is conserved. 
In this case the scaling relation (Eq.~\ref{eq:scaling}) becomes
\begin{equation}
\big(\frac{l_{\rm cl}}{l_{\rm driv}}\big) \rho_{\rm cl} \frac{v_{\rm cl}^{3}}{l_{\rm cl}} = \rho \frac{v_{\rm turb}^{3}}{l_{\rm driv}}\ .
\end{equation}
With $\Phi_{\rm A}= \Phi_{\rm V} H/r_{\rm cl} = \rho/\rho_{\rm cl} H/r_{\rm cl}$ one obtains
\begin{equation}
\frac{v_{\rm cl}^{3}}{l_{\rm cl}} = \Phi_{\rm A} \frac{v_{\rm turb}^{3}}{l_{\rm driv}}\ .
\end{equation}

\begin{figure}
        \resizebox{\hsize}{!}{\includegraphics[angle=-90]{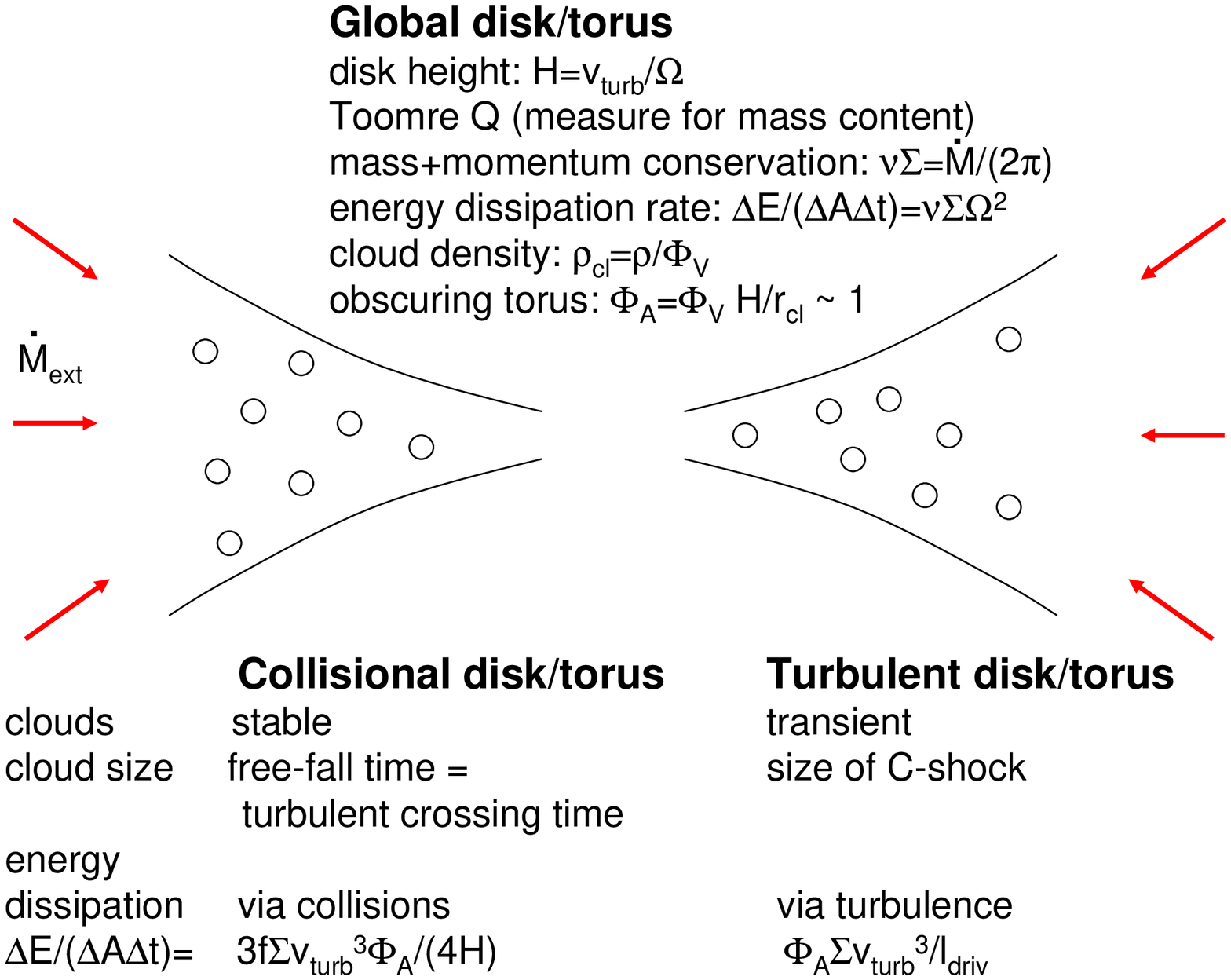}}
        \caption{Schematic of the collisional and turbulent disk/torus models. 
        } \label{fig:sketch}
\end{figure}

If the energy dissipation rate at large-scales is dominated by the energy dissipation rate in the dense clouds, 
\begin{equation}
\frac{\Delta E}{\Delta A \Delta t}=\Phi_{\rm A} \Sigma_{\rm cl} \frac{v_{\rm cl}^{3}}{l_{\rm cl}}=\Sigma \frac{v_{\rm cl}^{3}}{l_{\rm cl}}=
\Phi_{\rm A} \Sigma \frac{v_{\rm turb}^{3}}{l_{\rm driv}}\ .
\label{eq:eatturb}
\end{equation}
Since the viscous energy dissipation rate of the disk is given by 
\begin{equation}
\frac{\Delta E}{\Delta A \Delta t}=\nu \Sigma \Omega^{2}\ ,
\label{eq:nusig}
\end{equation}
the effective turbulent viscosity is
\begin{equation}
\nu = \Phi_{\rm A} v_{\rm turb} H
\label{eq:nuphia}
\end{equation}
for $D=2$. If $D=3$, it becomes $\nu = \Phi_{\rm V} v_{\rm turb} H$.
The disk mass accretion rate $\dot{M}$ can be estimated using
mass conservation (see, e.g., Pringle 1981)
\begin{equation}
\nu \Sigma = \frac{\dot{M}}{2 \pi}\ .
\end{equation}
In general, the viscosity is expected to be of the form
\begin{equation}
\nu= \zeta v_{\rm turb} H\ {\rm with}\ \Phi_{\rm V} \leq \zeta \leq 1\ .
\label{eq:nuturb}
\end{equation}
We will estimate $\zeta$ in Sect.~\ref{sec:gc}. Fig.~\ref{fig:sketch} visualizes the differences between the collisional and turbulent
clumpy disk models.

\section{Turbulent versus collisional disks \label{sec:versus}}

Within the collisional model the assumption of a constant cloud mass and sound speed leads to a constant density, size, and
surface density of the clouds independent of the distance to the galactic center (Fig.~\ref{fig:sketch}):
$r_{\rm cl} \propto c_{\rm s}^{-2}$ (Eq.~\ref{eq:rcoll}), $\rho_{\rm cl} \propto c_{\rm s}^{6}$ (Eq.~\ref{eq:rhocoll}), 
$N_{\rm cl} \propto c_{\rm s}^{4}$ (Eq.~\ref{eq:Ncoll}).
On the other hand, the area and volume filling factors depend on the sound speed and the angular velocity:
$\Phi_{\rm A} \propto \Omega v_{\rm turb}/ (Q c_{\rm s}^{4})$ (Eq.~\ref{eq:phiacoll}), $\Phi_{\rm V} \propto \Omega^{2}/(Q c_{\rm s}^{6})$
(Eq.~\ref{eq:phivcoll}).
In the case of a constant sound speed the obscuration of the nucleus is mainly due to a higher area filling factor of
clouds of same mass, density, and size at the inner edge of the torus where the area filling factor $\Phi_{\rm A}$ is maximum.
Thus clouds at the inner edge obscure the nucleus with the highest probability.

The situation is different in the turbulent model, where the volume and area filling factors are constant (Eq.~\ref{eq:phivturb}, \ref{eq:phiaturb}, 
and \ref{eq:Aphiaturb}). Thus, clouds located at all distance to the galactic center obscure the nucleus with the same probability
at optical wavelengths.
On the other hand, the cloud size and mass vary with the inverse of the angular velocity: 
$l_{\rm cl} \propto \Omega^{-1}$ (Eq.~\ref{eq:lturb} and \ref{eq:Alturb}), $M_{\rm cl} \propto \Omega^{-1}$ (Eq.~\ref{eq:mcl} and \ref{eq:Amcl}). 
For the cloud density and surface density we found
$\rho_{\rm cl} = \Phi_{\rm V}^{-1} \rho \propto \Omega^2$ and $N_{\rm cl} \propto \Omega$ (Eq.~\ref{eq:Nturb} and \ref{eq:ANcl}). 
For a constant rotation velocity $\Omega \propto R^{-1}$.
Thus, clouds at the inner edge of the torus have the highest surface densities. If an AGN torus is Compton thick
in X-ray emission, the optical thickness is most probably provided by a cloud located close to the central engine.

Another difference between the turbulent and collisional models lies in the turbulent viscosity and thus the derived disk mass accretion rate.
The effective viscosity of the collisional disk is $\nu_{\rm coll} \sim 3/4\,f \Phi_{\rm A} v_{\rm turb} H$ (Eq.~\ref{eq:tau}),
that of the turbulent disk is $\nu_{\rm turb}= \zeta v_{\rm turb} H$ with $\Phi_{\rm V} \leq \zeta \leq 1$ (Eq.~\ref{eq:nuturb}).
For $D=2$, the turbulent viscosity and the associated mass accretion rate is a factor of $\sim 1/f \sim 10$ higher than the
effective viscosity of a collisional disk.

Are the collisional and turbulent disk models exclusive? It is remarkable that the CND clouds have densities close to stability
limit against tidal shear and can be modeled by Bonner-Ebert spheres with sound or Alfv{\'e}nic speeds of $\sim 1$~km\,s$^{-1}$
(Vollmer et al. 2001a; 2001b). We can only speculate that intermittent turbulence knows about stability, in the sense that
if the kinetic energy of a turbulent clump with a velocity dispersion of $6$-$7$~km\,s$^{-1}$ is rapidly dissipated, the
resulting cloud with an Alfv{\'e}nic speed of $\sim 1$~km\,s$^{-1}$ will be marginally stable.

\section{Application of the models \label{sec:application}}

The collisional model (Sect.~\ref{sec:collisional}) was successfully applied to the circumnuclear disk (Vollmer et al. 2004) and
thick obscuring tori in nearby AGNs (Vollmer et al. 2008). We thus concentrate on the turbulent model (Sect.~\ref{sec:turbulent})
and its relation to the collisional model.

\subsection{The circumnuclear disk in the Galactic Center \label{sec:gc}}

The central black hole of the Galaxy is surrounded by a clumpy structure of gas and dust forming a disk or a ring ({\it
Circumnuclear Disk CND}, see e.g. G\"{u}sten et al. 1987) up to a radius of $\sim$5 pc\footnote{We assume $8.5$~kpc for the distance
to the Galactic Center.}. The CND has a total gas mass of a few
times 10$^{4}$~M$_{\odot}$ (Etxaluze et al. 2011, Requena-Torres et al. 2012) which is distributed in clouds with an
estimated area filling factor of $\phi_{\rm A} \sim 0.1$ and a volume filling factor of $\phi_{\rm V} \sim 0.01$. The clouds have
densities of $\sim 10^{5-6}$~cm$^{-3}$, sizes of $\sim 0.1$~pc and gas temperatures $\ga 200$~K. A typical dusty cloud has a
visual extinction A$_{\rm V} > 30^{\rm m}$ and M$_{\rm cl} \sim30$~M$_{\odot}$. Further properties of the clouds in the
central 2~pc are listed in Jackson et al. (1993). The vertical thickness ($2\,H$) of the CND increases from $\sim$0.5~pc at a
radius of 2~pc to about 2~pc at the outer radius of 7~pc. The disk rotates with a velocity of $\sim 110$~km\,s$^{-1}$ which
corresponds to a dynamical mass of $\sim 5 \times 10^{6}$~M$_{\odot}$ at the inner edge. The velocity dispersion 
($0.4$ times the linewidth) of the clumpy gas disk is $\sim$ 20~km\,s$^{-1}$ (G\"{u}sten et al. 1987), that of the clouds 
is $\sim 7-15$~km\,s$^{-1}$ (Montero-Casta{\~n}o et al. 2009). The observed properties of the CND and its gas clouds
are  summarized in Table~\ref{tab:cnd}.

For the moment, the CND represents the only gas torus for which we can estimate the turbulent viscosity via
its luminosity-to-mass ratio. 
The model luminosity-to-mass ratio due to turbulent viscous energy dissipation is (Eq.~\ref{eq:nusig} and \ref{eq:nuturb}): 
\begin{equation}
\begin{array}{ll}
L/M= & \nu \Omega^{2} = \zeta v_{\rm turb}H \Omega^{2} = \\
 & 7.3\,\zeta \big(\frac{v_{\rm turb}}{20~{\rm km\,s}}\big)^{2} \big(\frac{\Omega}{55~{\rm km\,s^{-1}pc^{-1}}}\big)~{\rm erg\,cm^{-2}s^{-1}}\ .
\end{array}
\label{eq:lm}
\end{equation}
Since the temperature of the molecular gas is high ($\ga 200$~K), the luminosity
due to turbulent viscous energy dissipation is dominated by O{\sc i}, H$_{2}$, H$_2$O, and high-level CO emission.
Submillimeter continuum emission can be used to determine the total gas mass of the CND.
Bradford et al. (2005) observed the CND up to a radius of $\sim 2$~pc in the CO(7-6) rotational transition. The combination with
lower CO rotational transitions  yielded temperatures of $T \sim 200-300$~K and a total mass of warm gas of at least $2000$~M$_{\odot}$. 
The derived luminosity-to-mass ratio is $L/M \sim 2.7$~L$_{\odot}$M$_{\odot}^{-1} = 5.3$~erg\,s$^{-1}$g$^{-1}$
taking into account CO, H$_{2}$, H$_2$O, and O{\sc i} emission of the warm gas. 
Given that the total gas mass of the CND is about a factor of $5-10$ higher than the estimated mass of warm gas (Etxaluze et al. 2011, Requena-Torres et al. 2012),
the total luminosity-to-mass ratio of the CND is $L/M \sim 0.5-1$~erg\,s$^{-1}$g$^{-1}$.
By inserting this value into Eq.~\ref{eq:lm}, we derive $\zeta \sim 0.1$. 
This is consistent with $D=2$, for which $\zeta = \Phi_{\rm A} \sim 0.1$ (Eq.~\ref{eq:nuphia}).
We thus suggest that intermittency leads to a fractal dimension of $D=2$ and lowers the turbulent viscosity of accretion-driven clumpy gas tori by one 
tenth or the area filling factor. 

In the following, we derive the scaling of the density and turbulent velocity dispersion with cloud size.
We have seen in Sect.~\ref{sec:turbulent} that for $D=2$ the scaling is $v_{l} \propto l^{\alpha}$ if $\rho_{l} \propto l^{-3\alpha}$.
The exponent $\alpha$ can thus be determined with the ratios $\rho/\rho_{\rm cl}=\Phi_{\rm V}$ and $2H/l_{\rm cl}$.
We set the disk thickness to be $2H=2v_{\rm turb} \Omega^{-1}=0.7$~pc,
the total gas mass of the CND within a radius of $2$~pc to be $M_{\rm CND}=10^{4}$~M$_{\odot}$, and thus $Q \sim (v_{\rm turb}/v_{\rm rot})
(M_{\rm dyn}/M_{\rm CND})=100$. With $\rho=\Omega^2/(\pi G Q)$, the cloud size is $l_{\rm cl}=0.1$~pc (Eq.~\ref{eq:clumpsize}).
With a density ratio between the two scales $l_{\rm cl}$ and $2H$ of $\rho_{\rm cl}/\rho=\Phi_{\rm V}^{-1}=28$,  
we derive $\alpha=0.56$. This exponent is very close to that found for galactic
molecular clouds ($\alpha=0.5$; Larson 1981, Solomon et al. 1987; however, 
Heyer et al. 2009 found a $\Sigma^{\frac{1}{2}}$ dependence of $v_{l}/l^{\frac{1}{2}}$).
The turbulent velocity thus scales with $v_{l} \propto l^{0.5}$ and the turbulent velocity of the clouds is
$v_{\rm turb}^{\rm cl} = 7$~km\,s$^{-1}$. This is compatible with, but at the lower end of the observed velocity dispersion (Montero-Casta{\~n}o et al. 2009).

According to Eq.~\ref{eq:phivturb}, \ref{eq:mcl}, and \ref{eq:Nturb}, the mass of a model gas cloud is $M_{\rm cl} \sim 30$~M$_{\odot}$, the column density is $N_{\rm cl} \sim 3 \times 10^{23}$~cm$^{-2}$, 
and the density is $n_{\rm cl} \sim 10^{6}$~cm$^{-3}$.
This density is consistent with the estimates based on HCN lines (Jackson et al. 1993), but at odds with cloud densities derived from
multi-transition CO observations (Requena-Torres et al. 2012). The area filling factor of the clouds is $\Phi_{\rm A}=0.26$ (Eq.~\ref{eq:phiaturb}).
Since these clouds are not selfgravitating, the Virial theorem cannot be applied to derive their masses. 
With the Virial cloud masses of Christopher et al. (2005) the CND would have a gas mass and a mass accretion rate of an AGN torus.
The mass accretion rate of the CND (Eq.~\ref{eq:nuturb} with $\zeta=\Phi_{\rm A}$) is $\dot{M}=10^{-2}$~M$_{\odot}$yr$^{-1}$. 
The assumed disk parameters and calculated cloud parameters for the turbulent disk model are summarized in Table~\ref{tab:cnd}.
In comparison to this, we now turn to the collisional disk model.

\begin{table}
      \caption{Observed/assumed CND and derived cloud properties within the turbulent clumpy disk model.}
         \label{tab:cnd}
      \[
       \begin{array}{lllll}
	  & & {\rm obs} & {\rm model} & \\
	 \hline
	 {\rm Alfven\ velocity}^{({\rm a})} & v_{\rm A} & 1 & & {\rm km\,s}^{-1} \\
	\hline
	{\rm disk\ radius} & R & 2 & & {\rm pc} \\
	{\rm CND\ gas\ mass} & M_{\rm CND} & 10^{4} & &  {\rm M}_{\odot} \\
	{\rm rotation\ velocity} & v_{\rm rot} & 110 & & {\rm km\,s}^{-1} \\
        %\hline
	{\rm CND\ velocity\ dispersion} & v_{\rm turb} & 20 & &  {\rm km\,s}^{-1} \\
	%\hline
	{\rm Toomre\ parameter} & Q & 100 & & \\
	%\hline
	%\hline
	{\rm disk\ height} & H & 0.36 & & {\rm pc} \\
	\hline
	{\rm mass\ accretion\ rate}^{({\rm b})} & \dot{M} & & 1{\rm -}10 \times 10^{-3} & {\rm M}_{\odot}{\rm yr}^{-1} \\
	\hline
	{\rm cloud} & & & & \\
	{\rm area\ filling\ factor} & \Phi_{\rm A} & 0.1 & 0.26 & \\
	{\rm volume\ filling\ factor} & \Phi_{\rm V} & 0.01 & 0.035 & \\
	{\rm mass} & M_{\rm cl} & 30 & 30 & {\rm M}_{\odot} \\
	{\rm size} & l_{\rm cl} & 0.1 & 0.1 & {\rm pc} \\
	{\rm velocity\ dispersion} & v_{\rm cl} & 7-15 & 7 & {\rm km\,s}^{-1} \\
	{\rm density} & \rho_{\rm cl} & 10^{5-6} & 10^{6} & {\rm cm}^{-3} \\
	{\rm column\ density} & N_{\rm cl} & 0.3{\rm -}3 & 3  & 10^{23}\ {\rm cm}^{-2} \\
	\hline
       \end{array}
      \]
\begin{list}{}{}
 \item[$^{\rm{(a)}}$] The Alfv\'{e}n velocity is assumed to be constant over all scales.
   \item[$^{\rm{(b)}}$] Derived quantities. The range corresponds to $\nu = \Phi_{\rm V} v_{\rm turb} H$  and  $\nu = \Phi_{\rm A} v_{\rm turb} H$.
 \end{list}
\end{table}

In this framework of a collisional disk, the CND has an area filling factor of $\Phi_{\rm A}=4/3 \vartheta^{-1} \simeq 0.1$ ($\vartheta=15$; Vollmer et al. 2004)
within a radius of $7$~pc. According to Eq.~\ref{eq:cloudmass}, the area filling factor at a radius of $R=2$~pc has to increase to $\Phi_{\rm A}=0.35$ to 
obtain a cloud mass of $M_{\rm cl}=10$~M$_{\odot}$. The cloud radius, density, and column density are then $r_{\rm cl}=0.03$~pc, 
$\rho_{\rm cl}=1.3 \times 10^{6}$~cm$^{-3}$, and $N_{\rm cl}=1.3 \times 10^{23}$~cm$^{-2}$. The volume filling factor at this radius is $\Phi_{\rm V}=0.03$.

The turbulent and collisional models thus give similar cloud properties, except for the cloud velocity dispersion, which is $1$~km\,s$^{-1}$ in the
collisional model and $7$~km\,s$^{-1}$ in the turbulent model. Since in the collisional
model the clouds are marginally selfgravitating by definition, the clouds of the turbulent model are highly overpressured 
with respect to the turbulent pressure at a given mass. 
A further difference between the turbulent and collisional disks is the mass accretion rate. The collisional model yields a $\sim 10$ times
smaller mass accretion rate (Vollmer et al. 2004) than the turbulent model.

\subsection{Clumpy gas tori in active galactic nuclei \label{sec:clumpytorus}}

Obscuring gas tori in local AGNs have a high velocity dispersion ($v_{\rm turb} \sim 40-80$~km\,s$^{-1}$) and are
therefore thick (Davies et al. 2007). For a representative AGN torus, we chose $v_{\rm turb}=50$~km\,s$^{-1}$.
This corresponds to the maximum shock velocity which allows C-type shocks at a mean density of $n=10^{5}$~cm$^{-3}$ (Le Bourlot et al 2002).
The typical total gas mass is a few $10^{6}$~M$_{\odot}$ within a radius of $\sim 10$~pc (Hicks et al. 2009) and the Toomre parameter is $Q \sim 5$ 
(Vollmer et al. 2008). The torus height at a radius of $5$~pc is $H=v_{\rm turb}\Omega^{-1}=2.3$~pc.
The area filling factor of these clouds (Eq.~\ref{eq:phiaturb}) is $\Phi_{\rm A} = 0.7$, i.e. the torus is obscuring, as it is expected. 
The volume filling factor of the clouds (Eq.~\ref{eq:phivturb}) is $\Phi_{\rm V}=0.01$, comparable to that of the CND.

The assumed torus parameters and calculated cloud parameters are summarized in Table~\ref{tab:tori}.
The mass accretion rate of the torus (Eq.~\ref{eq:nuturb} with $\zeta=\Phi_{\rm A}$) is $\dot{M}=4.8$~M$_{\odot}$yr$^{-1}$. 
Thus, the mass accretion rate of a turbulent, thick, obscuring gas tori is a factor of $\sim 500$ higher than that of a turbulent CND.

The cloud mass and size at a radius of $5$~pc are $M_{\rm cl}\sim 100$~M$_{\odot}$ and
$l_{\rm cl}=0.08$~pc. The cloud density and column density are $n_{\rm cl} = 9 \times 10^{6}$~cm$^{-3}$ and $N_{\rm cl}=2 \times 10^{24}$~cm$^{-2}$.
These clouds are thus Compton-thick. 
The velocity dispersion of the clouds is $v_{\rm cl} \sim v_{\rm turb}\,(l_{\rm cl}/(2H))^{0.5} \sim  6$~km\,s$^{-1}$.
The torus clouds are more massive, smaller, have an about ten times higher density and surface density,
but, most importantly, have a comparable velocity dispersion compared to the CND clouds.

\begin{table}
      \caption{Observed/assumed torus and derived cloud properties within the turbulent clumpy disk model.}
         \label{tab:tori}
      \[
       \begin{array}{llll}
	 \hline
	 {\rm Alfven\ velocity}^{({\rm a})} & v_{\rm A} & 1 & {\rm km\,s}^{-1} \\
	\hline
	{\rm torus\ radius} & R & 5 & {\rm pc} \\
	{\rm torus\ gas\ mass} & M_{\rm torus} & 1.2 \times 10^{6} & {\rm M}_{\odot} \\
	{\rm rotation\ velocity} & v_{\rm rot} & 110 & {\rm km\,s}^{-1} \\
        %\hline
	{\rm torus\ velocity\ dispersion} & v_{\rm turb} & 50 & {\rm km\,s}^{-1} \\
	%\hline
	{\rm Toomre\ parameter} & Q & 5 & \\
	%\hline
	%\hline
	{\rm torus\ height} & H & 2.27 & {\rm pc} \\
	\hline	
	{\rm mass\ accretion\ rate}^{({\rm b})} & \dot{M} & 0.2{\rm -}4.8 & {\rm M}_{\odot}{\rm yr}^{-1} \\ 
	\hline
	{\rm cloud^{({\rm c})}\ area\ filling\ factor} & \Phi_{\rm A} & 0.41 & \\
	{\rm cloud\ volume\ filling\ factor} & \Phi_{\rm V} & 0.014 & \\
	{\rm cloud\ mass} & M_{\rm cl} & 115 & {\rm M}_{\odot} \\
	{\rm cloud\ size} & l_{\rm cl} & 0.08 & {\rm pc} \\
	{\rm cloud\ velocity\ dispersion} & v_{\rm cl} & 6 & {\rm km\,s}^{-1} \\
	{\rm cloud\ density} & \rho_{\rm cl} & 9 \times 10^{6} & {\rm cm}^{-3} \\
	{\rm cloud\ column\ density} & N_{\rm cl} & 2 \times 10^{24} & {\rm cm}^{-2} \\
	\hline
       \end{array}
      \]
\begin{list}{}{}
 \item[$^{\rm{(a)}}$] The Alfv\'{e}n velocity is assumed to be constant over all scales.
 \item[$^{\rm{(b)}}$] Derived quantity. The range corresponds to $\nu = \Phi_{\rm V} v_{\rm turb} H$  and  $\nu = \Phi_{\rm A} v_{\rm turb} H$.
% \item[$^{\rm{(c)}}$] Values in brackets are derived from the model assuming a frozen-in magnetic field instead of recombination-ionization equilibrium (see Appendix A).
 \item[$^{\rm{(c)}}$] All cloud properties are derived from the model.
 \end{list}
\end{table}

Since typical radial extents of obscuring tori in nearby AGNs are about $10$~pc (Davies et al. 2007, Lira et al. 2013, 
M\"{u}ller-Sanchez et al. 2013), the typical cloud sizes and masses have
to be multiplied by a factor $2$ (both quantities depend on $\Omega^{-1}$): $M_{\rm cl} \sim 230$~M$_{\odot}$, $l_{\rm cl} \sim 0.16$~pc.
The cloud density has to be divided by a factor $4$ and the column density by a factor $2$: $\rho_{\rm cl}=2 \times 10^{6}$~cm$^{-3}$ and
$N_{\rm cl}=10^{24}$~cm$^{-2}$.

As for the CND (Sect.~\ref{sec:gc}), the cloud properties of the turbulent model are very similar to those of the collisional model, except
for the cloud velocity dispersion, which is $\sim 5$ times higher in the turbulent model compared to the collisional model. 
The derived mass accretion rate of the turbulent torus is about a factor of $10$ higher than the mass accretion rate 
for collisional tori (Vollmer et al. 2008). This is mainly due to the the mean fraction of cloud mass participating
in the highly supersonic cloud collisions $f=0.1$ (Eq.~\ref{eq:eatcoll}), which enters the viscosity prescription in the collisional model.

\section{The quenching of star formation in accretion-driven turbulent gas disks \label{sec:quench}}

As shown in Sect.~\ref{sec:application}, the main difference between the gas cloud in the collisional and turbulent clumpy
disk models lies in the velocity dispersion of the clouds. Whereas it is $\sim 1-1.5$~km\,s$^{-1}$ in the collisional model, it
is $6-7$~km\,s$^{-1}$ in the turbulent model. By definition, the clouds in the collisional model are marginally selfgravitating, i.e.
their free fall time equals the turbulent crossing time: $t_{\rm ff} = t_{\rm turb}$ (Eq.~\ref{eq:tfftturb}).
To assess the role of selfgravitation of the clouds, we calculate the fraction 
\begin{equation}
\frac{t_{\rm ff}}{t_{\rm turb}}=\sqrt{\frac{3 \pi \Phi_{\rm V}}{32 G \rho}}\frac{v_{\rm turb,cl}}{r_{\rm cl}}\ ,
\label{eq:tratio}
\end{equation}
where $v_{\rm turb,cl}$ is the turbulent velocity dispersion of the cloud. A cloud is selfgravitating if $t_{\rm ff}/t_{\rm turb} < 1$. 

Inserting Eqs.~\ref{eq:phivturb}, \ref{eq:lturb}, $r_{\rm cl}=l/2$, and $\rho=\Omega^{2}/(\pi G Q)$ into Eq.~\ref{eq:tratio} yields
\begin{equation}
\frac{t_{\rm ff}}{t_{\rm turb}}=13 \frac{v_{\rm turb,cl}}{v_{\rm turb}}= 13 \sqrt{\frac{l_{\rm cl}}{2H}}\ .
\label{eq:tfftturb1}
\end{equation}
We recall that the disk height is $H=v_{\rm turb}/\Omega$.
The cloud properties of Tables~\ref{tab:cnd} and \ref{tab:tori} lead to $t_{\rm ff}/t_{\rm turb} \sim 5$ for the CND and
$t_{\rm ff}/t_{\rm turb} \sim 2$ for a thick, obscuring torus. 
Since our value of the velocity dispersion of CND gas clouds is at the lower limit of the observed velocity dispersion
(Montero-Casta{\~n}o et al. 2009), the values of $t_{\rm ff}/t_{\rm turb}$ can also be seen as lower limits.
 
Thus, for both types of clouds selfgravitation does not play a role, i.e.
these clouds are overpressured with respect to the turbulent pressure and do not collapse to form stars. The high turbulent energy input caused by external mass accretion
thus leads to overpressured gas clouds with high velocity dispersions which cannot form stars. 
The accretion-driven turbulence thus suppresses or quenches star formation in the clumpy disk.
We suggest that the increase of the turbulent velocity in the torus gas is due to adiabatic compression of the torus gas by material
that falls onto the torus from outside. 

We suggest two ways to estimate the necessary external mass accretion rate onto the torus, which leads to an enhanced
gas velocity dispersion and thus an enhanced Toomre $Q$ parameter of the large-scale, smeared-out gas disk (with a mean density and a mean velocity 
dispersion\footnote{For $Q > 1$ the Toomre parameter corresponds to a measure of the gas content with $Q=1$ for the maximum gas content 
(see Sect.~\ref{sec:theory}).}): 
\begin{enumerate}
\item
Following Robertson \& Goldreich (2012; Eq.~6) the necessary condition for adiabatic compression of the torus gas is that the
compression timescale has to be smaller than the turbulent timescale times the efficiency of the energy cascade $\eta$: 
\begin{equation}
t_{\rm comp} = \frac{\lambda}{v_{\rm comp}} < \eta\ t_{\rm turb} = \eta \frac{H}{v_{\rm turb}}\ ,
\label{eq:compress}
\end{equation}
where $\lambda$ is the characteristic compression lengthscale.
Assuming $\lambda=H$ for convenience and taking into account intermittency ($\eta=\zeta$), 
this condition translates into $v_{\rm comp} > \zeta v_{\rm turb}$
with $\Phi_{\rm V} \leq \zeta \leq 1$ (see Eq.~\ref{eq:nuturb}).
The compression velocity can be calculated from the conservation of the rate of momentum transfer:
$\rho_{\rm infall} v_{\rm infall}^{2} = \rho_{\rm torus} v_{\rm comp}^{2} > \zeta \rho_{\rm torus} v_{\rm turb}^{2}$. It is assumed that
the gas radially hits the outer border of the torus which has an area of $2 \pi R H$. The external mass accretion rate is
\begin{equation}
\dot{M}_{\rm ext}= 2 \pi R H \rho_{\rm infall} v_{\rm infall} > \zeta \rho_{\rm torus} \frac{v_{\rm turb}^{2}}{v_{\rm infall}} 2 \pi R H\ .
\end{equation}
With $v_{\rm turb}/v_{\rm rot}=H/R$, $\Sigma_{\rm torus}=\rho_{\rm torus}H$, and $v_{\rm infall}=v_{\rm rot}$ one obtains
\begin{equation}
\dot{M}_{\rm ext} > 2 \pi \zeta v_{\rm turb} H \Sigma_{\rm torus} = \dot{M}.
\end{equation}
The external mass accretion rate has thus to be higher than the disk/torus mass accretion rate.
This is expected, because the torus is fed by the external mass infall. Based on the results of Sect.~\ref{sec:clumpytorus}
the external mass accretion rate is expected to be of the order of a few solar masses per year.
\item
the energy supply due to accretion $\Delta E/(\Delta A \Delta t) \sim \dot{M}\Omega^{2}$, where $\dot{M}$ is the disk/torus
mass accretion rate, has to be larger than the energy supply from supernovae (Vollmer \& Beckert 2003) $\Delta E/(\Delta A \Delta t) \sim 
\xi \dot{\Sigma}_{*} \sim v_{\rm rot}^{2} \dot{\Sigma}_{*} \sim \dot{M}_{*}\Omega^{2}$. The star formation rate can be
estimated via $\dot{M}_{*} \sim 0.02 M_{\rm gas} \Omega$ (Genzel et al. 2010, Daddi et al. 2010). We thus find $\dot{M} >  0.02 M_{\rm gas} \Omega$.
For $v_{\rm rot}=100$~km\,s$^{-1}$, $R=10$~pc, and $M_{\rm gas}= 5 \times 10^{6}$~M$_{\odot}$, the lower limit
for the mass accretion rate is $\dot{M} = 1$~M$_{\odot}$yr$^{-1}$. 
\end{enumerate}

The impact of AGN feedback on gas expulsion and driving of turbulence is not considered, because it is beyond the scope of this work. 
We conclude that continuous external mass accretion rate of a few solar masses per year leads to large-scale adiabatic compression of the torus gas
which increases the mean velocity dispersion and quenches star formation.
At small-scales, it is likely that adiabatic compression 
due to infall will be followed by adiabatic expansion, especially in turbulent flows. Here we only consider the overall behavior of the whole gas disk
in the presence of continuous external mass accretion. If the external mass accretion is time-dependent and stochastic(e.g., Gaspari et al. 2013), 
the initial massive gas infall makes the disk/torus thick. In the presence of a sufficient external mass accretion rate at the outer edge averaged over
a viscous timescale, the torus can sustain its thickness during a viscous timescale
\begin{equation}
t_{\rm visc}=R^{2}/\nu=\Omega^{-1} \big(\frac{v_{\rm rot}}{v_{\rm turb}}\big)^{2} \Phi_{\rm V}^{-1} \sim 5~{\rm Myr}\ .
\label{eq:tvisc}
\end{equation}
For the model to be valid, the variation timescale of the external mass accretion rate thus should not exceed the viscous timescale of the disk/torus. 

Once the external mass accretion has significantly decreased, the turbulent energy is dissipated in a crossing time
$H/v_{\rm turb}=\Omega^{-1}$, the disk becomes thinner and begins to form stars. In the case of a massive AGN torus, the subsequent 
star formation will most likely be qualified as a nuclear starburst.

Vollmer et al. (2008) proposed an evolutionary model which is divided into three phases:
Phase I - initial massive infall and formation of a turbulent, massive gas disk:
an initial infall of a large amount of gas , $M_{\rm gas} \sim 10^{6}~M_{\odot}$ leads to the formation of a massive ($Q \sim 1$), moderately thin 
($v_{\rm turb}/v_{\rm rot} < 5$) gas disk in which star formation proceeds.
Phase II - torus evolution at constant turbulent velocity:
during the first phase of its evolution the massive collisional torus stays thick. This implies that the mass accretion rate within the torus, and 
thus also the external mass accretion rate $\dot{M}$, do not decrease significantly during this phase.
Phase III - torus evolution at constant gas mass:
once the external mass accretion rate has significantly decreased, the torus evolves at constant gas mass.

The evolution of a thick, obscuring torus is proposed to proceed in the following way:
due to an initial, massive infall a massive ($Q \sim 1$) turbulent star-forming disk is formed (Phase I). The turbulence in this disk is 
maintained through the energy supply by feedback from rapid star formation. The subsequent SN explosions destroy the disk structure after 10 Myr, 
i.e. the intercloud medium is removed leaving only the densest, most massive clouds which remain Jeans-stable. The disk becomes collisional and 
stays geometrically thick (Phase II). After $\sim$100 Myr the mass accretion rate decreases and the disk becomes thin (Phase III) and ultimately transparent.

Based on the present work, we suggest an alternative evolution scenario including the three phases of Vollmer et al. (2008)
which depends on the external gas infall timescale $t_{\rm infall}=M_{\rm gas}/\dot{M}$. For $t_{\rm infall} \la 100 \Omega^{-1}$ the 
torus becomes thick and star formation is quenched.
\begin{itemize}
\item
Phase I
a torus with a small amount of gas ($\sim 10^{4}$~M$_{\odot}$) becomes thick for $\dot{M} \ga 10^{-3}$~M$_{\odot}$yr$^{-1}$.
In this phase gas is added to the torus without a significant gas consumption by star formation.
\item
Phase II
once the torus has acquired a gas mass exceeding $M_{\rm gas} \ga M_{\rm dyn} (10{\rm\ km\,s}^{-1}/v_{\rm rot}) \sim 10^{6}$~M$_{\odot}$ and 
$t_{\rm infall} \ga 100 \Omega^{-1}$, star formation sets in with $\dot{M}_{*} \sim 0.02 M_{\rm gas} \Omega$.
Since the infall is slow, the star formation rate exceeds the mass accretion rate ($\dot{M}_{*} > \dot{M}$), and
turbulence is maintained through the energy supply by feedback from rapid star formation. 
The subsequent SN explosions might destroy the disk structure after 10 Myr, removing the intercloud medium.
Otherwise, star formation consumes the gas within a timescale of $t_{\rm *} = M_{\rm gas}/\dot{M}_{*}$.
\item
Phase III - if the mass accretion rate increases, exceeding the expected star formation rate according to
$\dot{M}_{*} \sim 0.02 \times M_{\rm gas} \Omega$ (implying $t_{\rm infall} \la 100 \Omega^{-1}$), the torus gas is adiabatically compressed.
This leads to an increase of the velocity dispersion and the Toomre $Q$ parameter.
The turbulent energy supply of the torus gas becomes dominated by the gain of potential energy via gas accretion.
The turbulent torus clouds become overpressured with respect to the turbulent pressure and cannot collapse anymore. Therefore, star formation is
quenched in the turbulent, thick, obscuring torus. The subsequent high mass accretion rate powers the central AGN. 
\end{itemize}
If the external mass accretion rate decreases significantly, star formation sets
in again, producing a nuclear starburst (Phase II). The torus looses gas via star formation, SN gas expulsion, and
radial gas accretion within the torus. The necessary mass accretion rate to form a thick torus thus varies significantly with time
depending on the torus gas mass $(\dot{M} \ga 3 \times 10^{-3} M_{\rm gas} \Omega$).
The thickness and star formation rate of the torus entirely depend on the external mass accretion rate.

\section{Comparison with NGC~3227 and NGC~1068 \label{sec:seyfert}}

The dense molecular gas in active galactic nuclei generally has temperatures in excess of $100$~K.
Krips et al. (2011) analyzed various lower CO, HCN and HCO+ transitions observed at $< 2'' \sim 140$~pc resolution in NGC~1068.
They concluded that the gas is at temperatures $>200$~K. Hailey-Dunsheath et al. (2012) detected far-IR CO rotational emission from
the central $10'' \sim 700$~pc of NGC~1068. The detected transitions were modeled as arising from two different components at $170$ and $570$~K
with densities exceeding $4 \times 10^{5}$~cm$^{-3}$.

Davies et al. (2012) presented an LVG analysis of subarcsecond-resolution observations of CO(2--1) and HCN(1--0)
line emissions in the central region of the Seyfert galaxy NGC~3227. Based on the high HCN(1--0)/CO(2--1) ratio in the central 
80~pc of NGC~3227 and the central regions of NGC~6951 and NGC~1068, Davies et al. (2012) suggested that the dense gas
in all three galaxies is warm ($T \sim 300$~K), dense ($n_{\rm cl} \sim 10^{5.5}$~cm$^{-3}$, and have large velocity gradients
(${\rm d}V/{\rm d}r \sim 100$~km\,s$^{-1}$pc$^{-1}$)\footnote{However, it cannot be excluded that the HCN emission
is optically thick. In this case, the gas density and temperature cannot be derived from line ratio.}. 
Assuming a cloud size of $0.1$~pc, the clouds have a turbulent velocity dispersion of $10$~km\,s$^{-1}$.

Within our turbulent torus model (Sect.~\ref{sec:clumpytorus}), the derived cloud density ($\rho_{\rm cl}=\Omega^{2}/(\pi G Q)$) 
corresponds to an outer torus radius of $R \sim 30$~pc. The clumps at this radius have sizes between $0.3$ and $0.5$~pc and
masses between $180$ and $720$~M$_{\odot}$. Since the cloud size $l_{\rm cl}$ and the torus height $H$ are both
proportional to the angular velocity $\Omega$, the turbulent velocity of the clouds is the same for all radii $R$:
$v_{\rm cl}= 5$-$7$~km\,s$^{-1}$. These cloud properties are consistent with those derived by Davies et al. (2012).
All tori clouds are overpressured with respect to the turbulent pressure expected for selfgravitating clouds (Eq.~\ref{eq:tfftturb1}).
To estimate the temperature of gas in the model clouds, we follow Bradford et al. (2005) and require equilibrium 
between turbulent dissipation and (H$_{2}$, H$_{2}$O, CO)-line cooling within the turbulent clouds:
\begin{equation}
\Lambda_{\rm heat}= \rho_{\rm cl} \frac{v_{\rm cl}^{3}}{l_{\rm cl}} = \Lambda_{\rm cool}\ .
\end{equation}
For the cooling we use the model of Neufeld \& Kaufman (1993) with CO and H$_{2}$O abundances of $10^{-4}$ and $10^{-6}$, respectively.
With a cloud density of $n_{\rm cl}=3 \times 10^{5}$~cm$^{-3}$, a cloud size of $0.3$~pc, and a turbulent velocity dispersion
of $v_{\rm cl}=7$~km\,s$^{-1}$, we obtain a gas temperature of  $T=170$~K. Half of the cooling is provided by $H_{2}$ line-emission,
$40$\,\% by CO line-emission, and the rest by $H_{2}$O line-emission.
To reach a gas temperature of $300$~K a velocity dispersion of $v_{\rm cl}=14$~km\,s$^{-1}$ is required.
As already suggested in Sect.~\ref{sec:gc} for the CND, our model cloud velocity dispersion might be seen as a lower limit.

M\"{u}ller-Sanchez et al. (2009) presented subarcsecond-resolution H$_{2}$ $2.12$~$\mu$m line observations of the inner $300$~pc
of the Seyfert galaxy NGC~1068. They found two prominent linear structures leading to the active galactic nucleus from the north and south. 
The kinematics of the gas in these features are dominated by noncircular motions and indicate that material streams toward the nucleus on 
highly elliptical or parabolic trajectories. These gas streamers provide the central $10$~pc with fresh gas, corresponding to the
external mass accretion rate discussed in Sect.~\ref{sec:quench}. From their data, M\"{u}ller-Sanchez et al. (2009) derive an upper limit
for this external mass accretion rate of $\dot{M}_{\rm ext} \sim 15$~M$_{\odot}$yr$^{-1}$. This limit is entirely consistent with
the theoretical limit for external mass accretion rate of a few solar masses per year which leads to adiabatic compression and 
subsequent quenching of the star formation rate (Sect.~\ref{sec:quench}). The absence of rotation in the H$_{2}$ velocity field 
of the inner $20$~pc implies that the torus is probably hidden by the southern gas streamer (M\"{u}ller-Sanchez et al. 2009). This
leads to a torus size of $\sim 7$~pc, comparable to that of the CND. We thus suggest that the nucleus is in a Phase III, where the
torus is still formed through massive external accretion.

\section{Implications for star formation in the Galactic Center \label{sec:sfrgc}}

The quenched star formation in gas with a high turbulent velocity ($Q > 1$) might also help to elucidate the
paradox of youth in the Galactic Center (Ghez et al. 2003, Genzel et al. 2010). The central parsec of the Galaxy contains $\sim 200$ young, massive stars, 
and is one of the richest and massive star formation regions in the entire Galaxy. This is highly surprising. The presence of so many young stars 
in the immediate vicinity of the central massive black hole is unexpected (Allen \& Sanders 1986, Morris 1993, Ghez et al. 2003, Alexander 2005). 
For gravitational collapse to occur in the presence of the tidal shear from the central mass, gas clouds have to be denser than the critical Roche density,
which is higher than $10^{10}$~cm$^{-3}$ at a distance of $0.1$~pc from the black hole.
A population of $100$ young, massive stars is arranged mainly in two coherent disk structures (Bartko et al. 2009).
Since the transport of massive stars from more benign sites of formation is not favored by current observations, in situ star formation is the most
probable explanation for the paradox of youth. 
Recent simulations suggest that rapid dissipation and cooling in a molecular cloud plunging into the vicinity of a massive black hole or in
two clouds colliding near near a massive black hole can plausibly 
overcome the critical Roche density at $\sim 0.1$ to $1$~pc (Bonnell \& Rice 2008, Hobbs \& Nayakshin 2009). 
The resulting clumpy, filamentary disk fragments and may form stars efficiently.

Quenched star formation due to strong shocks at high velocities in colliding gas filaments can help to bring a fairly large amount of gas ($\geq 10^{4}$~M$_{\odot}$)
near the central black hole without star formation occuring on its way to the center. This gas can then settle at a distance of
$0.1$~pc from the black hole without a previous considerable gas consumption by star formation. This might help to explain the lack of young massive stars at 
galactic radii $> 0.5$~pc and the relatively low eccentricities of these stars within the central parsec of the Galaxy (Bartko et al. 2009).

\section{Conclusions \label{sec:conclusions}}

Galactic gas-gas collisions involving a turbulent multiphase ISM share common properties: 
dense extraplanar gas visible in CO, large linewidths ($\ga 50$~km\,s$^{-1}$),
strong mid-infared H$_{2}$ line emission, low star formation activity, and strong radio continuum emission in the absence of a high
star formation rate. Gas-gas collisions can occur in the form of ram pressure stripping caused by the rapid motion of a spiral galaxy within the
intracluster medium (NGC~4438; Vollmer et al. 2005, 2009), a galaxy head-on collision (NGC~4438; Taffy galaxies, Condon et al. 1993,
Gao et al. 2003, Braine et al. 2003, Vollmer et al. 2012), or a compression of the intragroup gas and/or galaxy ISM by an intruder galaxy which flies through
the galaxy group at a high velocity (Stephan's Quintet; Appleton et al. 2006, Guillard et 2009, Cluver et al. 2010, Guillard et al. 2012).

Gas tori in active galactic nuclei share many properties with these systems. The gas which falls from larger galactic radii
onto an existing gas torus hits the torus gas at high relative velocities. Shocks form which give rise to gas turbulence within the torus.
Davies et al. (2007) showed that the gas tori of nearby AGNs have high velocity dispersions, i.e. they are thick, and that 
there have been recent, but no longer active, starbursts in the last $10$-$300$~Myr, i.e. the star formation has recently decreased
despite the large torus gas masses ($\sim 10^{6}$~M$_{\odot}$). 

Inspired by these similarities, we developed an analytical model for turbulent clumpy gas disks where the energy to drive turbulence is supplied
by external infall or the gain of potential energy by radial gas accretion within the disk (Sect.~\ref{sec:theory}).
The gas disk is assumed to be stationary ($\partial \Sigma/\partial t=0$) and the external mass accretion rate to be close to the 
mass accretion rate within the disk (the external mass accretion rate feeds the disk at its outer edge). 
The external and disk mass accretion rates averaged over the viscous timescale are assumed to be constant.
In reality, the external mass accretion rate is certainly time-dependent and stochastic (see, e.g. Gaspari et al. 2013).
Therefore, non-steady hydrodynamics should prevail in real gas tori. Our model might be applied when all time-dependent quantities
are averaged over a sufficiently long timescale, i.e. the viscous timescale of the disk or several Myr.
Within the model, the disk is characterized by the disk mass accretion rate $\dot{M}$ and the Toomre $Q$ parameter which is
used as a measure of the gas content of the disk.
We suggest that the velocity dispersion of the torus gas is increased through adiabatic compression by the infalling gas.

In contrast to our previous collisional and turbulent models, the gas clouds are not assumed to be selfgravitating. The disk velocity dispersion is
fixed by the mass accretion rate and the gas surface density (Eq.~\ref{eq:nuturb}) (via Q). Turbulence is assumed to be supersonic, creating
shocks in the weakly ionized dense molecular gas. For not too high shock velocities ($<50$~km\,s$^{-1}$) these shocks will be continuous (C-type).
The cloud size is determined by the size of a C-shock at a given velocity dispersion (Eq.~\ref{eq:clumpsize}). If the turbulent velocity
dispersion is much higher, strong dissociative J-type shocks will develop and our determination of the cloud size can no longer be applied.
We give expressions for the expected volume and
area filling factors, mass, density, column density, and velocity dispersion of the clouds. The latter is based on 
scaling relations of intermittent turbulence (Eq.~\ref{eq:scaling}) whose open parameters are estimated for the Circumnuclear Disk
in the Galactic Center.

The results for the turbulent disk model were compared to those of the collisional model developed in Vollmer et al. (2004, 2008).
In the collisional model with constant cloud mass, the cloud size, density, and column density only depend on the gas temperature,
whereas the cloud area and volume filling factors depend on $\Omega$ and $\Omega^{2}$, respectively. 
In the case of a constant sound speed the obscuration of the nucleus is mainly due to a higher area filling factor of
clouds of same mass, density, and size at the inner edge of the torus where the area filling factor $\Phi_{\rm A}$ is maximum.
Within the turbulent model, we found $l_{\rm cl} \propto \Omega^{-1}$, $M_{\rm cl} \propto \Omega^{-1}$,
$\rho_{\rm cl} = \Phi_{\rm V}^{-1} \rho \propto \Omega^2$, and $N_{\rm cl} \propto \Omega$. 
Thus, clouds at the inner edge of the torus have the highest surface densities. If an AGN torus is Compton thick
in X-ray emission, the optical thickness is most probably provided by a cloud located close to the central engine.
The mass accretion rates of the turbulent torus is expected to be a factor of $\sim 10$ higher than those of the collisional model.

The turbulent and collisional model were applied to the CND and thick obscuring tori in AGNs.
Despite the different gas masses (CND: $10^{4}$~M$_{\odot}$; AGN torus: $10^{6}$~M$_{\odot}$) and 
velocity dispersions (CND: $20$~km\,s$^{-1}$; AGN torus: $50$~km\,s$^{1}$),
both models share gas clouds of similar masses and sizes. However, whereas the expected FWHM linewidths are $\sim 2.5$~km\,s$^{-1}$
in the collisional model, they are $\sim 17$~km\,s$^{-1}$ in the turbulent model. 
Whereas the clouds are by definition selfgravitating in the collisional model, they are overpressured with respect to the turbulent pressure 
in the turbulent model.
This prevents gravitational collapse and star formation is significantly reduced in the torus. This is consistent with the
findings of Davies et al. (2007) in nearby AGNs and the large observed linewidths of CND clouds (Montero-Casta{\~n}o et al. 2009).
In particular, the model gas cloud properties and the external mass accretion rate necessary for the quenching
of the star formation rate due to adiabatic compression are consistent with those derived from
observations (M\"{u}ller-Sanchez et al. 2009; Davies et al. 2012).

With these ideas in mind, we propose the following scenario for the evolution of a galactic center gas torus:
a torus with a small amount of gas ($\sim 10^{4}$~M$_{\odot}$) becomes thick for $\dot{M} \ga 10^{-3}$~M$_{\odot}$yr$^{-1}$.
In this phase gas is added to the torus without a significant gas consumption by star formation.
Once the torus has acquired a gas mass exceeding $M_{\rm gas} \ga M_{\rm dyn} (10{\rm\ km\,s}^{-1}/v_{\rm rot}) \sim 10^{6}$~M$_{\odot}$ and 
$t_{\rm infall} \ga 100 \Omega^{-1}$, star formation sets in with $\dot{M}_{*} \sim 0.02 M_{\rm gas} \Omega$.
Since the infall is slow, the star formation rate exceeds the mass accretion rate ($\dot{M}_{*} > \dot{M}$), and
turbulence is maintained through the energy supply by feedback from rapid star formation. 
The subsequent SN explosions might destroy the disk structure after 10 Myr, removing the intercloud medium.
Otherwise, star formation consumes the gas within a timescale of $t_{\rm *} = M_{\rm gas}/\dot{M}_{*}$.
If the mass accretion rate increases above the expected star formation rate $\dot{M}_{*} \sim 0.02 \times M_{\rm gas} \Omega$
within a timescale smaller than the star formation timescale ($t_{\rm infall} \la 100 \Omega^{-1}$), the torus gas is adiabatically compressed.
This leads to an increase of the velocity dispersion and the Toomre $Q$ parameter.
The turbulent energy supply of the torus gas becomes dominated by the gain of potential energy via gas accretion.
The turbulent torus clouds become overpressured with respect to the turbulent pressure and cannot collapse anymore. Therefore, star formation is
quenched in the turbulent, thick, obscuring torus. The subsequent high mass accretion rate powers the central AGN. 
If the external mass accretion rate decreases significantly, star formation sets
in again, producing a nuclear starburst. The necessary mass accretion rate to form a thick torus thus varies significantly with time
depending on the torus gas mass. The thickness and star formation rate of the torus entirely depend on the external mass accretion rate.

We suggest that quenched star formation due to strong shocks at high velocities in colliding gas streams (a gas cloud falling onto a
pre-existing torus) and subsequent adiabatic compression can help to bring a fairly large amount of gas ($\geq 10^{4}$~M$_{\odot}$) from a torus
near the central black hole without star formation occuring on its way to the center. This gas can then settle around at a distance of
$0.1$~pc from the black hole without a previous considerable gas consumption by star formation. This might help to explain the 
paradox of youth (Ghez et al. 2003), i.e. the lack of young massive stars at 
galactic radii $> 0.5$~pc and the relatively low eccentricities of these stars within the central parsec of the Galaxy (Bartko et al. 2009).

%---------------------------------------------------------------

\appendix

\section{Frozen-in magnetic field}

If the so-called frozen-in condition ($n_{i}v_{i}=const$, where $n_{i}$ and $v_{i}$
are the ionized gas density and velocity; e.g. Wardle 1990) is used instead of recombination-ionization equilibrium Eq.~\ref{eq:lll} becomes 
\begin{equation}
l=\sqrt{2} \frac{v_{\rm A,0}}{\alpha x_{i} \rho}\ .
\end{equation}
With $\gamma=600$~cm$^{-\frac{3}{2}}$s$^{\frac{1}{2}}$ and $\zeta_{\rm CR}=10^{-15}$~s$^{-1}$ the size of the C-shock is
\begin{equation}
l=0.01 \frac{(v_{\rm A,0}/1~{\rm km\,s}^{-1})}{\sqrt{(n_{\rm H}/5 \times 10^{4}~{\rm cm}^{-3})}}~{\rm pc}\ .
\label{eq:Alturb}
\end{equation}
The area filling factor and cloud mass then become
\begin{equation}
\Phi_{A}= \frac{20}{\sqrt{Q}}\ .
\label{eq:Aphiaturb}
\end{equation}
For the cloud mass and surface density we find
\begin{equation}
\begin{array}{l}
M_{\rm cl}=0.03\,(\Omega/1.5 \times 10^{-5}~{\rm yr}^{-1})^{-1} \times \\
(Q/5)^{\frac{1}{2}}\big(v_{\rm A,0}/1~{\rm km\,s}^{-1}\big)^{2}\big(v_{\rm turb}/30~{\rm km\,s}^{-1}\big)~{\rm M}_{\odot}\ , 
\label{eq:Amcl}
\end{array}
\end{equation}
\begin{equation}
\begin{array}{ll}
N_{\rm cl}= & 8 \times 10^{22}\,(\Omega/1.5 \times 10^{-5}~{\rm yr}^{-1}) \times \\
 & (Q/5)^{-\frac{1}{2}}\big(v_{\rm turb}/30~{\rm km\,s}^{-1}\big)~{\rm cm}^{-2}\ .
\end{array}
\label{eq:ANcl}
\end{equation}
The cloud mass and surface density thus increase linearly with the turbulent velocity of the disk, whereas
in the recombination-ionization equilibrium model $M_{\rm cl} \propto v_{\rm turb}^{\frac{5}{2}}$ and
$N_{\rm cl} \propto v_{\rm turb}^{\frac{3}{2}}$. The frozen-in magnetic field approximation yields much too
small cloud sizes and masses in the CND for the high cosmic ray ionization rate measured by Yusef-Zadeh et al. (2013)
in the inner $100$~pc of the Galaxy.


\begin{thebibliography}{}

\bibitem{a1} Alexander T., 2005, PhR, 419, 65

\bibitem{a2} Allen D.~A., Sanders R.~H., 1986, Natur, 319, 191

\bibitem{a3} Appleton P.~N., Xu, K. C., Reach, W., et al., 2006, ApJ, 639, L51 

\bibitem{a4} Balkovsky, E., \& Falkovich, G. G. 1998, Phys. Rev. E, 57, 1231

\bibitem{a5} Bartko H., Martins, F., Fritz, T. K., et al., 2009, ApJ, 697, 1741

\bibitem{a6} Bonnell I.~A., Rice W.~K.~M., 2008, Sci, 321, 1060

\bibitem{a7} Bradford C.~M., Stacey G.~J., Nikola T., et al., 2005, ApJ, 623, 866 

\bibitem{a8} Braine J., Davoust E., Zhu M., et al., 2003, A\&A, 408, L13 

\bibitem{a9} Chen C.-Y., Ostriker E.~C., 2012, ApJ, 744, 124 

\bibitem{a10} Christopher M.~H., Scoville N.~Z., Stolovy S.~R., Yun M.~S., 2005, ApJ, 622, 346 

\bibitem{a11} Cluver M.~E., Appleton, P. N., Boulanger, F., et al., 2010, ApJ, 710, 248

\bibitem{a12} Combes F., Dupraz C., Casoli F., Pagani L., 1988, A\&A, 203, L9

\bibitem{a13} Condon J.~J., Helou G., Sanders D.~B., Soifer B.~T., 1993, AJ, 105, 1730

\bibitem{a14} Crutcher R.~M., 1999, ApJ, 520, 706 

\bibitem{a15} Davies R.~I., M{\"u}ller S{\'a}nchez F., Genzel R., et al., 2007, ApJ, 671, 1388

\bibitem{a16} Davies R., Mark D., Sternberg A., 2012, A\&A, 537, A133 

\bibitem{a16a} Dobbs, C.~L., \& Pringle, J.~E.\ 2013, MNRAS, 1187 

\bibitem{a17} Draine B.~T., 1980, ApJ, 241, 1021 

\bibitem{a18} Draine B.~T., Roberge W.~G., Dalgarno A., 1983, ApJ, 264, 485

\bibitem{a18a} Draine, B.~T., \& McKee, C.~F.\ 1993, ARA\&A, 31, 373 

\bibitem{a19} Etxaluze M., Smith H.~A., Tolls V., Stark A.~A., Gonz{\'a}lez-Alfonso E., 2011, AJ, 142, 134

\bibitem{a20} Fleck R.~C., Jr., 1996, ApJ, 458, 739 

\bibitem{a20a} Fraternali, F. \& Binney, J.~J. 2008, MNRAS, 386, 935

\bibitem{a21} Frisch, U., Sulem, P.-L., Nelkin, M., 1978, J. Fluid Mech., 87, 719

\bibitem{a22} Gao Y., Zhu M., Seaquist E.~R., 2003, AJ, 126, 2171

\bibitem{a22a} Gaspari, M., Ruszkowski, M., \& Oh, S.~P.\ 2013, MNRAS, accepted for publication, arXiv:1301.3130 

\bibitem{a23} Genzel R., Eisenhauer F., Gillessen S., 2010, RvMP, 82, 3121

\bibitem{a24} Ghez A.~M., Duch{\^e}ne, G., Matthews, K., et al., 2003, ApJ, 586, L127 

\bibitem{a25} Guillard P., Boulanger F., Pineau Des For{\^e}ts G., Appleton P.~N., 2009, A\&A, 502, 515

\bibitem{a26} Guillard P., Boulanger, F., Pineau des Forêts, G., et al., 2012, ApJ, 749, 158

\bibitem{a27} G\"{u}sten R., Genzel R., Wright M.C.H. et al., 1987, ApJ 318, 124

\bibitem{a27a} Hailey-Dunsheath S., Sturm, E.; Fischer, J., et al., 2012, ApJ, 755, 57 

\bibitem{a28} Heyer M., Krawczyk C., Duval J., Jackson J.~M., 2009, ApJ, 699, 1092 

\bibitem{a29} Hicks E.~K.~S., Davies R.~I., Malkan M.~A., et al., 2009, ApJ, 696, 448

\bibitem{a30} Hobbs A., Nayakshin S., 2009, MNRAS, 394, 191 

%\bibitem{a31} Hobbs A., Power C., Nayakshin S., King A.~R., 2012, MNRAS, 421, 3443

\bibitem{a32} Hwang J.-S., Struck C., Renaud F., Appleton P.~N., 2012, MNRAS, 419, 1780

\bibitem{a33} Jackson J.~M., Geis N., Genzel R., et al., 1993, ApJ, 402, 173

\bibitem{a34} Kenney J.~D.~P., Rubin V.~C., Planesas P., Young J.~S., 1995, ApJ, 438, 135 

\bibitem{a34a} Krips M., Martin, S., Eckart, A., et al., 2011, ApJ, 736, 37

\bibitem{a35} Kritsuk A.~G., Norman M.~L., Padoan P., Wagner R., 2007, ApJ, 665, 416 

\bibitem{a36} Krolik J.~H., Begelman M.~C., 1988, ApJ, 329, 702 

\bibitem{a37} Larson R.~B., 1981, MNRAS, 194, 809 

\bibitem{a38} Le Bourlot J., Pineau des For{\^e}ts G., Flower D.~R., Cabrit S., 2002, MNRAS, 332, 985 

\bibitem{a38a} Leroy, A.K., Walter, F., Brinks, E., et al. 2008, AJ, 136, 2782

\bibitem{a38b} Lin, D.~N.~C., \& Pringle, J.~E.\ 1987, ApJ, 320, L87 

\bibitem{a38c} Lira, P., Videla, L., Wu, Y., et al.\ 2013, ApJ, 764, 159 

\bibitem{a39} McKee C.~F., Li P.~S., Klein R.~I., 2010, ApJ, 720, 1612

\bibitem{a39a} Marinacci, F., Binney, J., Fraternali, F., et al. 2010, MNRAS, 404, 1464

\bibitem{a40} Montero-Casta{\~n}o M., Herrnstein R.~M., Ho P.~T.~P., 2009, ApJ, 695, 1477

\bibitem{a41} Morris M., 1993, ApJ, 408, 496 

\bibitem{a42} M{\"u}ller S{\'a}nchez F., Davies R.~I., Genzel R., et al., 2009, ApJ, 691, 749

\bibitem{a42a} M{\"u}ller-S{\'a}nchez, F., Prieto, M.~A., Mezcua, M., et al.\ 2013, ApJ, 763, L1

\bibitem{a43} Neufeld D.~A., Kaufman M.~J., 1993, ApJ, 418, 263 

\bibitem{a43a} Parravano, A. 1987, A\&A, 172, 280

\bibitem{a44} Peterson B.~W., Appleton, P.~N., Helou, G., et al., 2012, ApJ, 751, 11

\bibitem{a45} Pringle J.~E., 1981, ARA\&A, 19, 137

\bibitem{a46} Renaud F., Appleton P.~N., Xu C.~K., 2010, ApJ, 724, 80 

\bibitem{a47} Requena-Torres M.~A., G\"{u}sten, R., Wei{\ss}, A., et al., 2012, A\&A, 542, L21 

\bibitem{a47a} Robertson B., Goldreich, P., 2012, ApJ, 750, L31 

\bibitem{a48} Sani E., Davies, R. I., Sternberg, A., et al., 2012, MNRAS, 424, 1963 

\bibitem{a49} Solomon P.~M., Rivolo A.~R., Barrett J., Yahil A., 1987, ApJ, 319, 730 

\bibitem{a49a} Strasser, S.~T., Dickey, J.~M., Taylor, A.~R., et al., 2007, AJ, 134, 2252 

\bibitem{a50} Trinchieri G., Sulentic J., Breitschwerdt D., Pietsch W., 2003, A\&A, 401, 173 

\bibitem{a51} van der Hulst J.~M., Rots A.~H., 1981, AJ, 86, 1775

\bibitem{a52} Vollmer B. \& Duschl W.J. 2001a, A\&A, 377, 1016

\bibitem{a52a} Vollmer B., Duschl W.~J., 2001b, A\&A, 377, 1016

\bibitem{a53} Vollmer B. \& Beckert T. 2002, A\&A, 382, 872

\bibitem{a54} Vollmer B. \& Beckert T. 2003, A\&A, 404, 21

\bibitem{a55} Vollmer B., Beckert, T., \& Duschl, W.J. 2004, A\&A, 413, 949

\bibitem{a56} Vollmer B., Braine J., Combes F., Sofue Y., 2005, A\&A, 441, 473

\bibitem{a57} Vollmer B., Beckert T., Davies R.~I., 2008, A\&A, 491, 441

\bibitem{a58} Vollmer B., Soida M., Chung A., Chemin L., Braine J., Boselli A., Beck R., 2009, A\&A, 496, 669

\bibitem{a59} Vollmer B., Braine J., Soida M., 2012, A\&A, 547, A39

\bibitem{a60} Wardle, M. 1990, MNRAS, 246, 98

\bibitem{a61} Williams J.~P., Bergin E.~A., Caselli P., Myers P.~C., Plume R., 1998, ApJ, 503, 689 

\bibitem{a62} Yusef-Zadeh F., Hewitt, J. W., Wardle, M., et al., 2013, ApJ, 762, 33 

\end{thebibliography}
\end{document}